\documentclass[aps,
prd,10pt,
superscriptaddress]{revtex4-2}

\usepackage{makecell} 
\usepackage{graphicx}
\usepackage{dcolumn}
\usepackage{bm}
\usepackage[hidelinks]{hyperref} 
\usepackage{xcolor}
\usepackage{array} 
\usepackage{booktabs}
\usepackage{subfig} 

\usepackage{enumitem}
\usepackage{soul}

\usepackage{amsmath,amssymb,amsfonts}
\usepackage{mathtools}  
\usepackage{bm}         
\usepackage{physics}    
\usepackage{tensor}     
\usepackage{braket}     
\usepackage{slashed}    
\usepackage{mathrsfs}   

\usepackage{amsthm}

\usepackage{graphicx}
\usepackage{subcaption} 
\usepackage{float}      
\usepackage{caption}

\begin{document}

\title{A Brief Review of Wormhole Cosmic Censorship}

\author{Leonel Bixano}
    \email{Contact author: leonel.delacruz@cinvestav.mx}
\author{I. A. Sarmiento-Alvarado}%
 \email{Contact author: ignacio.sarmiento@cinvestav.mx}
\author{Tonatiuh Matos}%
 \email{Contact author: tonatiuh.matos@cinvestav.mx}
\affiliation{Departamento de F\'{\i}sica, Centro de Investigaci\'on y de Estudios Avanzados del Intituto Politécnico Nacional, Av. Intituto Politécnico Nacional 2508, San Pedro Zacatenco, M\'exico 07360, CDMX.
}%

\date{\today}

\begin{abstract}
Spacetime singularities, in the sense that curvature invariants are infinite at some point or region, are thought to be impossible to observe, and must be hidden within an event horizon. This conjecture is called Cosmic Censorship (CC), and was formulated by Penrose. Here we review another type of CC where spacetime singularities are causally disconnected from the universe, because the throat of a wormhole ``sucks in'' the geodesics and prevents them from making contact with the singularity. In this work, we present a series of exact solutions to the Einstein--Maxwell--Dilaton equations that feature a ring singularity; that is, the curvature invariants are singular in this ring, but the ring is causally disconnected from the universe so that no geodesics can touch it. This extension of CC is called Wormhole Cosmic Censorship.
\end{abstract}

\maketitle


\section{Introduction}\label{sec1}
{Wormholes, explored in general relativity, were first addressed by Ellis and Bronnikov~\cite{Ellis:1973yv,Bronnikov:1973fh}, who provided solutions involving scalar fields without event horizons. Morris and Thorne~\cite{Morris:1988cz} later described traversable wormholes requiring exotic matter. Visser~\cite{Cita:LorentzianWormholes} expanded this by developing Lorentzian and ``thin-shell'' wormholes, reducing the need for exotic matter. Recent research includes the Einstein–Maxwell--Scalar Field theory~\cite{Lobo:2005us,Goulart:2017iko,Lazov:2017tjs}, offering solutions with a phantom scalar field. In~this work, we will present one that meets energy conditions without exotic matter (Dilaton scalar field). 
}

{On the other hand, spacetime singularities are one of the most intriguing objects we study in general relativity. These singularities are understood as points or regions where some curvature invariant is singular. We believe that these types of singularities must be hidden by some type of physical object to make it impossible to see them and establish causal contact with them, thus preventing any physical violations.
In 1964, Penrose in~\cite{Penrose:1964wq} showed that gravitational collapse could create spacetime singularities under certain conditions. Around the same period, Hawking studied cosmological singularities. In~1969, Penrose in~\cite{Penrose:1969pc} suggested that such singularities are hidden by event horizons, forming the basis of the \textit{Cosmic Censorship Conjecture (CCC)}, 
 which was later refined by Hawking and Ellis~\cite{Hawking:1966sx,Hawking:1966jv,Hawking:1967ju}. Penrose’s (CCC) and Matos’s \textit{Wormhole Cosmic Censorship Conjecture} both aim to maintain cosmic causal order by concealing singularities. Penrose's applies to black holes using event horizons, while Matos's applies to traversable wormholes using the wormhole's topology.
The WCCC extends the CCC principle to wormhole spacetimes, \textit{applicable across multiple exact solutions and theories}. WCCC ensures that traversable wormholes have singularities that are causally hidden by the wormhole structure. This makes wormholes safer and more physically possible, allowing traversal without encountering singularities.
We demonstrate that certain precise solutions of Einstein's field equations, which exhibit spacetime singularities, possess a common attribute: the singularities remain unobservable as the wormhole absorbs the geodesics, thereby inhibiting the spacetime singularity from establishing causal interaction with the universe. We conduct a re-examination of some of these exact solutions and establish that they are, indeed, causally isolated from the universe.}

{Visser and Poisson~\cite{Poisson:1995sv} describe a ``cut and paste'' method for thin-shell wormholes and assess their stability. Lobo~\cite{Lobo:2005us} examines the use of phantom energy to support throats, highlighting necessary violations of the NEC in regular geometries. Gao, Jafferis, and~Wall~\cite{Gao:2016bin} make AdS wormholes traversable via a quantum mechanism involving negative average null energy. Maldacena, Milekhin, and~Popov~\cite{Maldacena:2018gjk} maintain a wormhole in asymptotically flat 4D using Casimir energy, resulting in a large smooth throat without causal violations.
Altogether, thin-shell (classical), phantom (classical with \mbox{NEC < 0}), and~semiclassical/AdS-CFT wormholes  
achieve traversability without visible singularities, which places the WCCC well in the modern landscape. When a singularity exists (e.g., ring in rotating solutions), it must remain causally untouchable. If~singularities does not exist (previous cases), consistency is immediate, and \textit{there is no reason for censoring singularities.}}

We start by examining the Einstein--Maxwell Lagrangian coupled to a dilaton or ghost scalar field. Notable theories studied in this work include Einstein--Maxwell theories with an uncoupled scalar field, low-energy superstrings (S-S), Kaluza--Klein (K-K), or~Entanglement Relativity (E-R). Thus, we start with the~Lagrangian 
\begin{equation}\label{LagrangianoTesisUnidades}
    \mathfrak{L}=\sqrt{-g}\bigg(-\frac{1}{\mu_0 \sigma_0}R +\frac{1}{\mu_0 \sigma_0}2\epsilon_0 (\nabla \phi)^2 + \frac{1}{\mu_0}e^{-2 \alpha_0 \phi } F^2 \bigg),
\end{equation}
where $ \sigma_0=8\pi G/(\mu_0c^4)$, $c$ represents the speed of light, $G$ denotes the gravitational constant, and~$\mu_0$ is the vacuum permeability, $R$ is Ricci scalar, $\phi$ is scalar field, and~ $F^2$ is invariant generated by the Faraday tensor. 
Here the constant $\epsilon_0$ can only take two values---1 if $\phi$ is a dilatonic scalar field and $-1$ if it is a ghost scalar field---while the constant $\alpha_0$ defines the theory in question. 
Prominent theories in this field include $\alpha_0^2 = {0, 1, 3}$, corresponding to the Einstein--Maxwell theories with an uncoupled scalar field, low-energy S-S, and~K-K theories. Notably, this formulation also encapsulates emerging theories, such as the E-R (applications of this theory can be found in \cite{Minazzoli:2025gyw,Minazzoli:2024zwo,Minazzoli:2025nbi} ) where $\alpha_0^{\,\, 2}=1/12$.

Employing variational methods to derive the associated field equations, we obtain the following: 
\begin{subequations}\label{EcuacionesDeCampoOriginales}
\begin{equation}\label{Eq:Campo1}
    \nabla_\mu \left( e^{-2\alpha_0 \phi} F^{\mu \nu} \right)=0,
\end{equation}
\begin{equation}\label{Eq:Campo2}
    \epsilon_0 \nabla^2 \phi+\frac{\alpha_0}{2} \sigma_0 \left( e^{-2\alpha_0 \phi} F^{2} \right)=0,
\end{equation}
\begin{equation}\label{Eq:Campo3}
    R_{\mu \nu}=2\epsilon_0 \nabla_\mu \phi \nabla_\nu \phi  + 2 \sigma_0 e^{-2\alpha_0 \phi} \left( F_{\mu \sigma} \tensor{F}{_\nu}{^\sigma} -\frac{1}{4} g_{\mu \nu } F^2 \right),
\end{equation}
\end{subequations}

In the Section \ref{sec2}, 
 we will present the symmetries considered of the spacetime, and the~solutions, alongside their respective behavior of the metric function and the equations for the constraint~parameters. 

The Section \ref{sec3} will address the comprehensive physical analysis of solution \eqref{SolucionLambdaCombinada}, which encompasses all relevant information from solutions \eqref{SolucionLambda5} and~\eqref{SolucionLambda6}. 

The Section \ref{sec4} provides the Penrose diagram and the causal structure, thereby demonstrating the validity of these solutions and their global~hyperbolicity. 

In Section~\ref{section: flat subspaces}, we will introduce an analogous method for obtaining new solutions in higher dimensions, utilizing $\alpha_0^{\, \, 2}=3$, i.e.,~within the framework of Kaluza--Klein~Theories.

Lastly, we will present a novel solution utilizing the methodology outlined in Section~\ref{section: flat subspaces}, which is currently under~study.

In this paper, $\mathbf{M}_n$ denotes the set of all $n \times n$ real matrices, and~the set of symmetric matrices is denoted by $\mathbf{Sym}_n$.
The centralizer of a subset $\mathscr{A} \subset \mathbf{M}_n$, denoted by $\mathcal{C} ( \mathscr{A} )$, is defined as $\mathcal{C} ( \mathscr{A} ) = \{ M \in \mathbf{M}_n : M A = A M \text{ for all } A \in \mathscr{A} \}$.

\section{Analysed~Solutions}\label{sec2}

To solve field Equation~(\ref{EcuacionesDeCampoOriginales}), we employed a stationary and axially symmetric spacetime, characterized by the existence of two Killing vectors $\{ \partial_t,\partial_\varphi \}$. In~Weyl\linebreak $(\rho=\sqrt{(x^2+1)(1-y^2)},z=Lxy)$, and~spheroidal ($Lx=r-l_1,y=\cos{\theta}$) (where ($r,\theta$) corresponds to Boyer--Lindquist coordinates) coordinate systems, the~metric is successively represented by the following forms:
 \begin{subequations}\label{ds}
\begin{align}
    ds^2&=-f\left\{ d(ct)-\omega d \varphi \right\}^2 +f^{-1} \left\{ e^{2k} (d\rho ^2 + dz^2) +\rho^2 d\varphi^2 \right\} \label{ds CilindricasDimensiones}\\
    &= -f\left( d(ct)-\omega d \varphi \right)^2  + f^{-1} \bigg( L^2(x^2+1)(1-y^2) d\varphi^2 \label{ds spheroidales} \\ & \qquad+L^2(x^2+y^2) e^{2k} \left\{ \frac{dx^2}{x^2+1} +\frac{dy^2}{1-y^2} \right\} \bigg), \notag
\end{align}
 \end{subequations}
where $\rho \in [0,\infty)$ ,$\{ z ,x \}\in \mathbb{R}$, $y \in [-1,1]$, $\theta \in [0,\pi]$, and~$r\in (-\infty,-l_1] \bigcup \ [l_1,\infty)$.

In the same way, considering these symetries, we can use the following anzat for the electromagnetic 4-potential:
\begin{equation}\label{4Potencial}
    A_{\mu}=\bigg[ A_t(\rho,z),0,0,A_\varphi (\rho,z) \bigg] .
\end{equation}

The family utilized to derive novel solutions is \textit{second class of solutions}:
\begin{equation}\label{SegundaClaseSoluciones}
    f=f_0, \quad \kappa=\kappa_0 e^{\lambda}, \quad \psi= \frac{\sqrt{f_0}}{\sqrt{\sigma_0} \kappa_0} e^{-\lambda} +\psi_0, \\ \quad \chi = \sqrt{\sigma_0} \sqrt{f_0}\kappa_0 e^{\lambda} + \chi_0, \quad \epsilon= b_0,
\end{equation} 
in which $\{f_0, \psi_0, \chi_0, \kappa_0, b_0 \}$ represent the integration constants, ($f,\epsilon,\psi ,\chi,\kappa$) denote the gravitational, rotational, electric, magnetic, and~scalar potentials analogous to Ernst potentials, and $\lambda(x,y)$ constitutes a general coordinate of a two-dimensional subspace that facilitates the solution of the field equations satisfying the Laplace equation.
\begin{equation}\label{Eq:LaplaceEnEsferoidales}
        \partial_x \{ (x^2+1)\partial_x \lambda \}+\partial_y \{(1-y^2) \partial_y \lambda \}=0.
\end{equation}

A comprehensive calculation and explanation to derive the corresponding family of solutions and beyond is provided in~\cite{Matos:2010pcd,Matos:2000ai}.

In~\cite{Matos:2000za}, two solutions to (\ref{Eq:LaplaceEnEsferoidales}) were introduced, referred to as $\lambda_5,\lambda_6$:
\begin{subequations}\label{LambdaSoluciones}
\begin{align}
       \lambda_5&=\lambda_0 \frac{x}{(x^2+y^2)} , \label{lambda5}\\
       \lambda_{6}&=\lambda_0 \frac{y}{(x^2+y^2)} , \label{lambda6}
   \end{align} 
\end{subequations}
where $\lambda_0$ is a integration~constant.

\subsection{Solution~5}

In~\cite{Bixano:2025jwm}, the~electromagnetic 4-potential and metric functions associated with $\lambda_5$ were derived and analyzed, and~are presented in the subsequent expressions:
\begin{subequations}\label{SolucionLambda5}
\begin{align}
        f&=f_0=1,\\
        \omega_5 &=-L\frac{ \lambda_0}{f_0 } \bigg( \frac{y(x^2+1)}{x^2+y^2} \bigg), \label{Omega Lambda5}\\
        A_{\varphi5} &=-\frac{\sqrt{f_0}}{2 \kappa_0 \sqrt{\sigma_0}} \frac{\omega_5}{L} e^{-\lambda_5}, \label{A3 Lambda5} \\
        A_{t5}&=\frac{\sqrt{f_0}}{2\kappa_0 \sqrt{\sigma_0}} e^{-\lambda_5}, \\
    k_{\lambda_5} &= -k_{0} \lambda_0 ^2  \frac{(1-y^2)}{4(x^2+y^2)^4} \bigg(  -8x^2y^2(x^2+1)   +(x^2+y^2)^2\big\{(1-y^2) +2(x^2+y^2)\big\} \bigg), \label{k Lambda5} 
\end{align}
\end{subequations}
where $k_0$ is a integration constant corresponding to the function metric $k(x,y)$.

\subsection{Solution~6}

The subsequent expressions present the electromagnetic 4-potential and metric functions related to $\lambda_6$, which were derived and examined in~\cite{DelAguila:2015isj}.
\begin{subequations}\label{SolucionLambda6}
\begin{align}
        f&=f_0=1, \\
        \omega_6 &=\frac{ L \lambda_0 }{f_0} \frac{x(1-y^{2})}{x^{2}+y^{2}}, \label{Omega Lambda6}\\
        A_{\varphi6} &=-\frac{\sqrt{f_0}}{2 \kappa_0 \sqrt{\sigma_0}} \frac{\omega_6}{L} e^{-\lambda_{6}}, \label{A3 Lambda6} \\
        A_{t6}&=\frac{\sqrt{f_0}}{2\kappa_0 \sqrt{\sigma_0}} e^{-\lambda_{6}} \\
    k_{\lambda_6} &= -k_{0} \lambda_0 ^2  \frac{(1-y^2)}{4(x^2+y^2)^4} \bigg(  8x^2y^2(x^2+1) -(x^2+y^2)^2(1-y^2)  \bigg) . \label{k Lambda6} 
\end{align}
\end{subequations}

\subsection{Combination~Solution}

Moreover,~\citet{Bixano:2025bio} analyzed the combination of (\ref{SolucionLambda5}) and (\ref{SolucionLambda6}), and~using this combination, we can encompass both previously mentioned~solutions.

It is important to note that, while the combination of the solutions is linear as per \mbox{term $\lambda$}
\begin{equation}\label{Lambda5+Lambda6}
    \lambda_c=\frac{\lambda_0 y + \tau_0 x}{(x^2+y^2)},
\end{equation}
the function metrics themselves are not. This is due to the necessity of resolving more complicate equations, as~elaborated in~\cite{Matos:2000za,Matos:2000ai,DelAguila:2015isj,Bixano:2025jwm,Bixano:2025bio}.

For the combination solution, the~functions metric and the electromagnetic\linebreak 4-potential are
\begin{subequations}\label{SolucionLambdaCombinada}
\begin{align}
        f&=f_0=1, \\
        \omega &=\frac{ L }{f_0} \bigg( \frac{\lambda_0 x(1-y^{2})-\tau_0 y (x^2+1)}{x^{2}+y^{2}} \bigg), \label{Omega LambdaCombinada}\\
        A_\varphi &=\frac{\sqrt{f_0}}{2 \kappa_0 \sqrt{\sigma_0}} \bigg( A_{3}-\frac{\omega}{L}e^{-\lambda_{c}} \bigg), \label{A3 LambdaCombinada} \\
        A_t&=\frac{\sqrt{f_0}}{2\kappa_0 \sqrt{\sigma_0}} \bigg( e^{-\lambda_{c}} -1\bigg) \label{A0 LambdaCombinada}\\
        k_{c} &= k_{\lambda5}+k_{\lambda6} -k_{0}\frac{8xy(1-y^2)(x^2+1)(x^2-y^2)\lambda_0 \tau_0}{4(x^2+y^2)^4}  \label{k LambdaCombinada},
\end{align}
\end{subequations}

For our objective, it is imperative to consider the asymptotic behavior of (\ref{SolucionLambdaCombinada}) 
\begin{subequations}\label{ComportamientoAsimptoticoOmega}
\setlength{\jot}{0pt}  
\begin{align}
        \lim\limits_{x \rightarrow \pm \infty } \omega(x,y) &= - \tau_0 L y, \label{OmegaxInfinito} \\
        \omega(0,y) &= - \frac{\tau_0 L}{y},\label{OmegaxCero} \\
        \omega(x,0) &= \frac{L \lambda_0}{x},\label{omegay0} \\
        \omega(x,1) &=-L \tau_0,\label{omegay1}
    \end{align}
\end{subequations}
\begin{subequations}\label{ComportamientoAsimptoticoK}
\begin{align}
        \text{For} \quad x \gg 1 \,\,\,\,\,\, \text{we have}\,\,\,\,\,\,  k_c \approx - k_0 \tau_0^2 \frac{(1-y^2)}{2x^2}, \label{kxInfinito} \\
        k_c(0,y)= \frac{k_0(1-y^2)}{4y^4} \big\{ \lambda_0^2 (1-y^2) -\tau_0^2 (1+y^2) \big\}, \label{kxCero} \\
        k_c (x,0)= k_0\frac{(\lambda_0^2-\tau_0^2(2x^2+1))}{
        4x^4}, \label{ky0} \\
        k_c (x,1)=0. \label{ky1}
    \end{align}
\end{subequations}

Using the solution provided in (\ref{SolucionLambdaCombinada}), which encapsulates all pertinent information related to (\ref{SolucionLambda5}) and (\ref{SolucionLambda6}), we shall exclusively include the Ricci scalar, Kretschmann scalar, and~the difference between density and pressure  (this expression is in the comoving frame, i.e.,~in the diagonal tetrad. For~additional details, refer to \cite{Morris:1988cz,DelAguila:2015isj,Bixano:2025jwm,Bixano:2025bio}) pertaining to (\ref{SolucionLambdaCombinada}).
\begin{subequations}
\begin{equation}
        R=\frac{(4k_0+1) e^{-2k_c}}{2L^2 (x^2+y^2)^4} \bigg\{ \lambda_0^2 \Big( y^2(1-y^2) +x^2(3y^2+1)  \Big) +\tau_0^2 \Big( y^2+x^2(x^2-3y^2+1) \Big) +4\lambda_0 \tau_0 x y (x^2-y^2) \bigg\}, \label{Ricci para lambdac}
    \end{equation}
\begin{align}
        KN=\frac{F_{c}(x,y)}{8L^{4}(x^2+y^2)^{12}} e^{-4k(x,y)}, \label{KN para lambdac}
    \end{align}
\end{subequations}
\begin{multline}\label{rho - varrho Lambdac}
    \varrho-P=\frac{ e^{-2k_c}}{2L^2 (x^2+y^2)^5}   \bigg\{ 4\lambda_0 \tau_0 x y( x^2-y^2 ) \Big( x^2+2y^2-1+4k_0(x^2+1) \Big)+ 2\lambda_0^2 \Big( y^4(1-y^2) +x^4(1+(1+8k_0)y^2 )
    \\ +2(4k_0+y^2) x^2 y^2 \Big) + \tau_0^2 \Big[ (4k_0+1)(x^2+1)x^4 -2x^2y^2 (4k_0(x^2+1)+x^2-3)+(1+4k_0 +(4k_0-7)x^2)y^4\Big]\bigg\},
\end{multline}
wherein $F_{c}(x,y)$ denotes a polynomial whose degree is inferior to $(x^2+y^2)^{12}$.


\subsection{Constraint~Parameters}

Substituting the solution (\ref{SolucionLambdaCombinada}) associated with $\lambda_c$, and~the scalar field
\begin{equation}\label{FormaDelPotencialSuma}
    \phi(x,y)=-\frac{\lambda_c}{\alpha_0},
\end{equation}
in (\ref{EcuacionesDeCampoOriginales}), the~constraint on the free parameters of the solution~is
\begin{equation}\label{EcuacionDeVerificacion}
    \alpha_0^2(4k_0+1)-4\epsilon_0=0 .
\end{equation}

Table \ref{TablaValoresk} summarises the results.

\begin{table}[b]
\caption{\label{TablaValoresk}%
{\color{blue} Values of $k_0$.}}
\begin{ruledtabular}
\begin{tabular}{ccc}
\textbf{$\alpha_0^2$} &
\textbf{Dilatonic Field ($\epsilon_0=1$)} &
\textbf{Phantom Field ($\epsilon_0=-1$)} \\
\colrule
1/12 & $47/4$ & $-49/4$ \\
1 & $3/4$ & $-5/4$ \\
3 & $1/12$ & $-7/12$ \\
4 & 0 & $-1/2$ \\
$n\geq 5$ & $(4-n)/4n$ & $-(4+n)/n$
\end{tabular}
\end{ruledtabular}
\end{table}

Upon careful consideration of (\ref{EcuacionDeVerificacion}), it becomes evident that $(4k_0+1)=4\epsilon_0 /\alpha_0^{\,\, 2}$, then
\begin{align*}
    &\text{if we have a Dilatonic field} \quad (\epsilon=1) \quad \Rightarrow \quad (4k_0+1)>0,\\
    &\text{if we have a Phantom field} \quad (\epsilon=-1) \quad \Rightarrow \quad (4k_0+1)<0.
\end{align*}

\section{Physical~Analysis}\label{sec3}

In this study, we present solely the most significant result. For~an in-depth exploration of the calculations, readers are referred to the detailed expositions in the papers~\cite{DelAguila:2015isj,Bixano:2025jwm,Bixano:2025bio,Bixano:2025qxp}.

\subsection{Singularities and Asymptotical~Flatness}

By employing Equations~(\ref{Ricci para lambdac}) and (\ref{KN para lambdac}), it is evident that a singularity exists at the point $x=y=0$, which is represented by \textit{ring singularity} $r=l_1$ and $\theta=\pi/2$ within the framework of Lewis--Papapetrou coordinates, or~alternatively by $\rho=L$ and $z=0$ in the context of Weyl~coordinates.

\begin{quote}
It is now understood that this singularity is causally disconnected; however, the~numerical analysis provided in~\cite{Matos:2012gj,DelAguila:2018gni,Bixano:2025bio} lays the groundwork for the early progress of the WCC. Subsequently, in~\cite{DelAguila:2023twe,Bixano:2025qxp}, this conjecture, analogous to the Cosmic Censorship proposed by Penrose~\cite{Penrose:1969pc} and further refined by Hawking and Ellis~\cite{Hawking:1973uf}, is demonstrated~analytically.

\end{quote}

Conversely, considering the limit as $r\rightarrow \pm \infty \Rightarrow x\rightarrow \pm \infty$, it becomes evident that both (\ref{Ricci para lambdac}) and (\ref{KN para lambdac}) converge to $0$, irrespective of the sign of $(4k_0+1)$. This indicates that the nature of a dilatonic or phantom scalar field is not important in terms of 
 asymptotic behaviour. The~solution \eqref{SolucionLambdaCombinada} presented in this work is \textit{asymptotically flat when we fix $\tau_0=0$}; consequently, the~solution analyzed in~\cite{DelAguila:2015isj} is the only one possessing physical significance; however, this does not imply that the combined solution lacks~importance.

\subsection{Null Energy~Conditions}

The characteristics of the expression (\ref{rho - varrho Lambdac}) are examined across various regions as follows:
\begin{subequations}\label{ComportamientoAsimptoticoRhoMenosP}
\begin{equation}\label{RhoMenosPInfinito}
        \text{For} \quad x \gg 1 \,\,\, \text{implies}\,\,\, (\varrho-P)\approx \frac{(4k_0+1) \tau_0^2}{2L^2x^4},
    \end{equation}
\begin{equation}\label{RhoMenosPXCero}
        \lim\limits_{x \rightarrow  0 } (\varrho-P)= \frac{e^{-2 k_c(0,y)}}{2 L^2 y^6} \Big\{ 2\lambda_0^2 (1-y^2) +\tau_0^2 (4k_0+1) \Big\},
    \end{equation}
\begin{equation}\label{RhoMenosPXY}
        \lim\limits_{x \rightarrow  y } (\varrho-P)= \frac{e^{-2 k_c(y,y)}}{16 L^2 y^6} \Big\{ 2\tau_0^2 (1-y^2)
        +\lambda_0^2 (4k_0+1) (1+y^2) \Big\},
    \end{equation}
\begin{equation}\label{RhoMenosPy0}
        (\varrho-P)_{|y=0}= \frac{e^{-2 k_c(x,0)}}{2 L^2 x^6} \Big\{ 2\lambda_0^2 +\tau_0^2 (4k_0 +1) (x^2+1)  \Big\},
    \end{equation}
\begin{equation}\label{RhoMenosPy1}
        (\varrho-P)_{|y=1}= (4k_0+1)\frac{(2\lambda_0 x+\tau_0 \{ x^2-1\})^2}{2L^2(x^2+1)^4}.
    \end{equation}
    
\end{subequations}

Upon conducting a thorough analysis of all the equations presented in (\ref{ComportamientoAsimptoticoRhoMenosP}), three significant observations can be discerned:

\begin{itemize}
    \item All the possible values of a exponential function is $e^{k} \in [0,\infty)$.
    \item Given $y\in [-1,1]$ and subsequently $y^{2n} \in [0,1] \quad $, this implies that $ \quad (1-y^2)>0$.
    \item The terms of $\lambda_0,\tau_0,x,y$ with even exponents yield solely positive values.
\end{itemize}

Therefore, the~crucial factor determining the fulfillment of the NEC is $(4k_0+1)$. In~other words
\begin{align*}
    &\textbf{Dilatonic Field} \quad \,\,\, \text{implies}\,\,\, \quad (\varrho-P)>0,  \\
    &\textbf{Phantom Field}  \quad \,\,\, \text{implies}\,\,\, \quad (\varrho-P)<0.
\end{align*}


\subsection{Geometries}
To determine the profile of the compact object or the form of the wormhole's throat, it is necessary to initially select a hypersurface with constants $t$ and $y$ of (\ref{ds}), and~subsequently embed it into a cylindrical plane space  (for more details on the method, see~\cite{Lobo:2017cay,DelAguila:2023twe,Bixano:2025bio}).

\begin{figure}[H]

\subfloat[\centering]{\includegraphics[width=7cm]{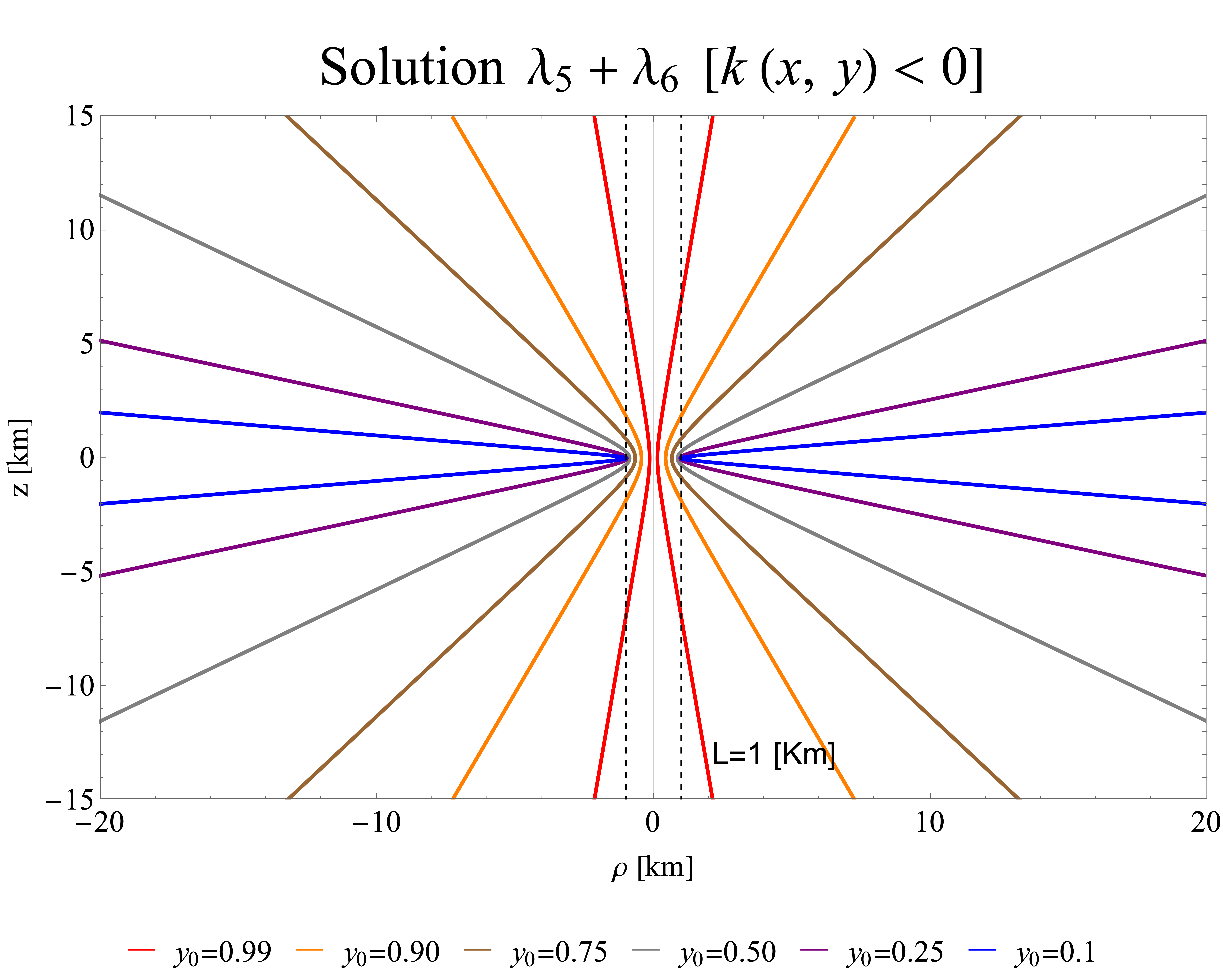}}
\subfloat[\centering]{\includegraphics[width=7cm]{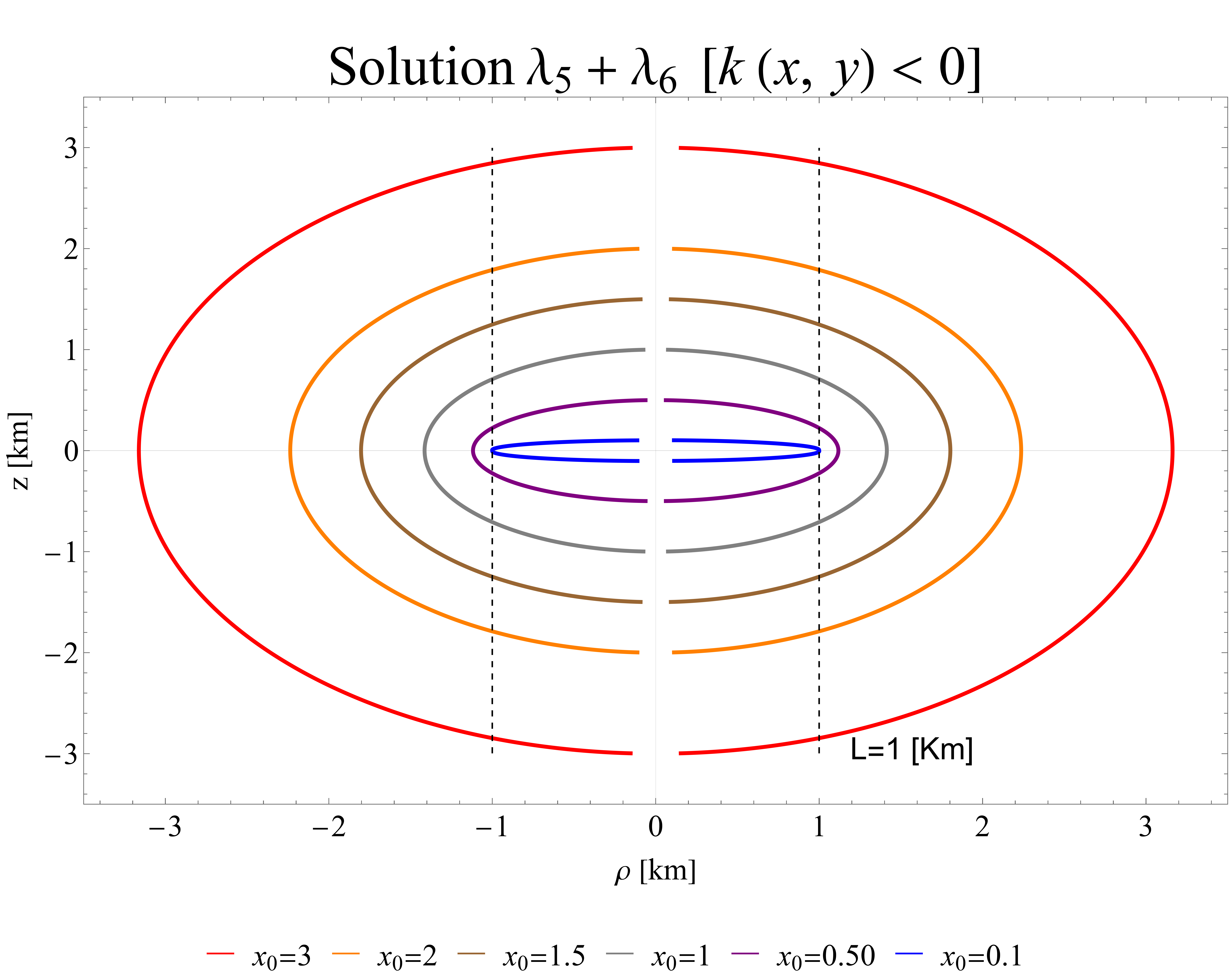}}\\
\subfloat[\centering]{\includegraphics[width=9cm]{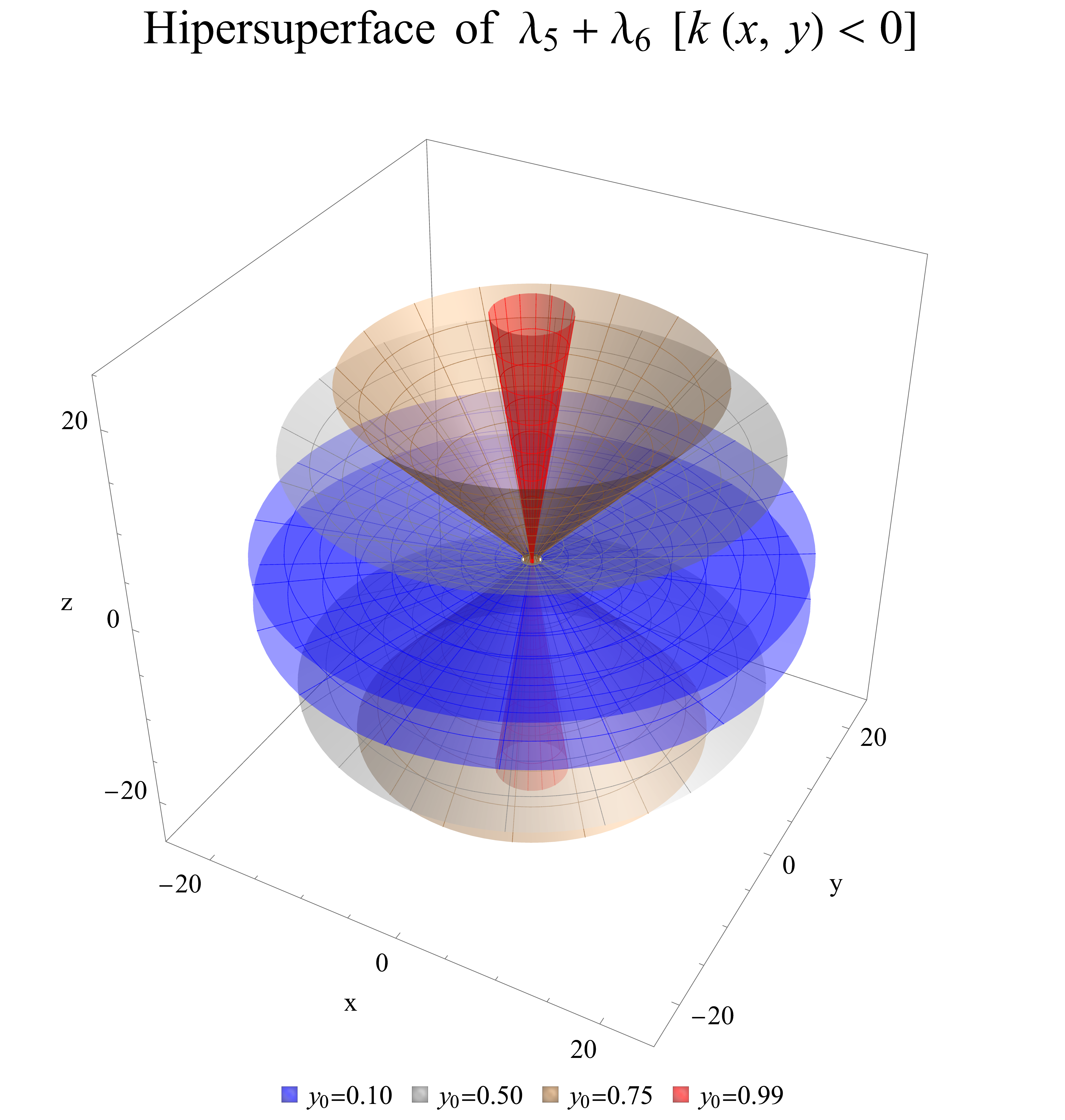}}

\caption{All units in the graphs are expressed in kilometers and were taken from~\cite{Bixano:2025bio}.
Embedding diagrams corresponding to solution (\ref{SolucionLambdaCombinada}): (\textbf{{a}}) Wormhole profile. (\textbf{{b}})  Throat shape. (\textbf{{c}}) Revolution surface of the wormhole profile.}\label{fig:GeometriasL5L6}
\end{figure}

The differential equations of the embedded diagrams, which were solved using numerical methods with the initial condition $z(0)=0$ and parameters $\lambda_0=10^{-2}$, $\tau_0=10^{-3}$, $k_0=1/12$, $L=1$~km, are
\begin{subequations}\label{Ec de la GoemtriaHipersup ycte}
\begin{align}
        &\overline{\rho}(x,y_0)^2= L^2 (x^2+1)(1-y_0^2) -\omega(x,y_0)^2  , \label{rho(x) ycte} \\
        &\left( \frac{d\, \overline{\rho}}{dx} \right)^2 +\left( \frac{dz}{dx} \right)^2= L^2\frac{(x^2+y_0^2)}{x^2+1} e^{2k(x,y_0)}. \label{EcDif rhoZ(x) ycte}
    \end{align}
\end{subequations}
\begin{subequations}\label{Ec de la GoemtriaHipersup xcte}
\begin{align}
        &\overline{\rho}(x_0,y)^2= L^2 (x_0^2+1)(1-y^2) -\omega(x_0,y)^2  , \label{rho(x) xcte} \\
        &\left( \frac{d\ \overline{\rho}}{dy} \right)^2 +\left( \frac{dz}{dy} \right)^2= L^2\frac{(x_0^2+y^2)}{1-y^2} e^{2k(x_0,y)}. \label{EcDif rhoZ(x) xcte}
    \end{align}
\end{subequations}
The embedded diagrams can be viewed in Figure~\ref{fig:GeometriasL5L6}. The~black dashed lines in \mbox{Figure~\ref{fig:GeometriasL5L6}a,b} represent the size of the wormhole's throat, while each colored line corresponds to a specific profile or shape defined by $y_0$ and $x_0$, respectively. In~the case of Figure~\ref{fig:GeometriasL5L6}a, $z>0$ corresponds to one universe, $z<0$ corresponds to another universe or possibly the same universe at different coordinates, and~$z(0)=0$ denotes the throat.
The initial two figures of Figure~\ref{fig:GeometriasL5L6} indicate that traversing a wormhole along the polar axis results in a closure of the throat, thereby demonstrating that the wormhole is effectively closed in this direction, while presenting its maximum dimension within the equatorial plane. (These assumptions are valid for (\ref{SolucionLambda5}), (\ref{SolucionLambda6}), and~(\ref{SolucionLambdaCombinada}).)

This embedding technique demonstrates the dependency of the throat size on the angle, specifically, the~function
\begin{equation}\label{GargantaAxysimetrica}
    \rho(x,y)= L \sqrt{(x^2+1)(1-y^2)}=\sqrt{(r-l_1)^2+L^2} \sin{\theta},
\end{equation}
represents an effective radial function for wormholes that lack spherical symmetry (independence of $\theta$).

\subsection{Tidal~Forces}

To examine tidal forces, we can employ the methodology outlined in~\cite{Morris:1988cz}, as~utilized in~\cite{DelAguila:2015isj,Bixano:2025bio,Bixano:2025jwm}. This approach leverages a transformation from the comoving frame to an astronaut frame to analyze the geodesic deviation equation. The~transformation provides the following representation of the aforementioned equation:
\begin{equation}\label{CondicionFuerzasDeMarea}
    |\tensor{R}{_{\hat{\mu} \hspace{0.2em} \hat{0} \hspace{0.2em} \hat{\mu} \hspace{0.2em} \hat{0} }}| \leq \frac{g_{Earth}}{2m * c^2}\approx (10^5\text{Km})^{-2}.
\end{equation}
where the subscript $\hat{\mu}$ denotes the component of the Ricci tensor within the astronaut's reference~frame.

The results provide a comprehensive understanding of this class of wormholes (\ref{SegundaClaseSoluciones}). The~most secure method for traversing the wormhole is in proximity to the polar axis, although~not precisely along it. Near~the equatorial plane, traversal is not feasible due to the presence of exceedingly hazardous regions, particularly when considering components $\tensor{R}{_{\hat{\mu} \hspace{0.2em} \hat{0} \hspace{0.2em} \hat{\mu} \hspace{0.2em} \hat{0} }}$ and $\hat{\mu}=1,3$. 

Considering the tidal forces, a~wormhole that optimizes safety possesses a size of
\begin{equation}\label{Tamaño de HW seguro}
    L\geq 10^{3}\ \text{km} \quad \text{which implies that} \quad M_{WH}\geq 10^{3}M_{\odot},
\end{equation}
where $M_{\odot}$ is a \textit{solar mass}.

\subsection{Geodesics}
The geodesics of these wormholes were derived using the Hamiltonian formulation, wherein the corresponding Hamiltonian is as follows:
\begin{equation}\label{Hamiltoniano xy}
    2\mathcal{H} = \frac{f \big( l_z -\omega E\big)^2}{L^2(x^2+1)(1-y^2)}-\frac{E^2}{f}
    +\frac{f e^{-2k}}{L^2(x^2+y^2)} \bigg\{ (x^2+1) p_x^2 + (1-y^2) p_y^2 \bigg\},
\end{equation}
where the motion constants~are
\begin{subequations}\label{ConstantesMovimiento}
\begin{equation}\label{Energia}
        -p_t= E=f(c\dot{t}-\omega \dot{\varphi}),
    \end{equation}
\begin{equation}\label{Momento Angular}
        p_\varphi= l_z=\frac{L^2 (x^2+1)(1-y^2)}{f}\dot{\varphi}+\omega E,
    \end{equation}
\end{subequations}
and the other momenta~are
\begin{subequations}\label{Momentos px py}
\begin{equation}\label{px}
        p_x= \frac{L^2(x^2+y^2) e^{2k}}{f(x^2+1)}\dot{x},
    \end{equation}
\begin{equation}\label{py}
        p_y= \frac{L^2(x^2+y^2) e^{2k}}{f(1-y^2)}\dot{y}.
    \end{equation}
\end{subequations}

From (\ref{ConstantesMovimiento}), it is possible to compute the proper angular momentum
\begin{equation}\label{Velocidad Angular}
     \frac{ \dot{\varphi} }{ \dot{t} } =\frac{d \varphi}{dt}=\frac{f^2 (l_z-\omega E) c}{L^2(x^2+1)(1-y^2)E +f^2 \omega (l_z-\omega E)}.
\end{equation}

Thus, employing the following initial values corresponding to a null geodesic, 

\begin{equation}\label{Condiciones de E lz}
    l_z=1, \qquad E=2,\qquad 
    x(0)=25, \qquad p_x(0)=-2, \qquad p_y(0)=0,
\end{equation}
and arbitrary $y(0)=y_0$, it is feasible to compute the equation of motion~numerically

\begin{equation}\label{Ecuaciones de Hamilton}
        \dot{x}^{\mu} \equiv \frac{\partial \mathcal{H}}{\partial p_{\mu}}, \qquad \qquad 
        \dot{p}_{\mu} \equiv -\frac{\partial \mathcal{H}}{\partial x^{\mu}}.
    \end{equation}

and extend the solution into pseudo-Cartesian coordinate space
\begin{equation}\label{pseudocartesian}
    X_c=\rho \cos{(\frac{d \varphi}{dt} t)} ,\quad 
    Y_c=\rho \sin{(\frac{d \varphi}{dt} t)} ,\quad
    Z_c=z.
\end{equation}

The parameters used were $\lambda_0=10^{-2}$, $\tau_0=10^{-3}$, $k_0=1/12$, and~a sun size WH
\begin{subequations}\label{SunSize WH}
\begin{equation}\label{MasaDelSol}
    M=M_{\odot} \approx 2\times 10^{30}\ \text{kg},
\end{equation}
\begin{equation}\label{Condicion L SunSize}
    L <2l_1=r_s^{Sun} \approx 3 \times 10^3\  \text{m}.
\end{equation}
\end{subequations}

\begin{figure}[H]
\centering

\subfloat[\centering]{\includegraphics[width=7cm]{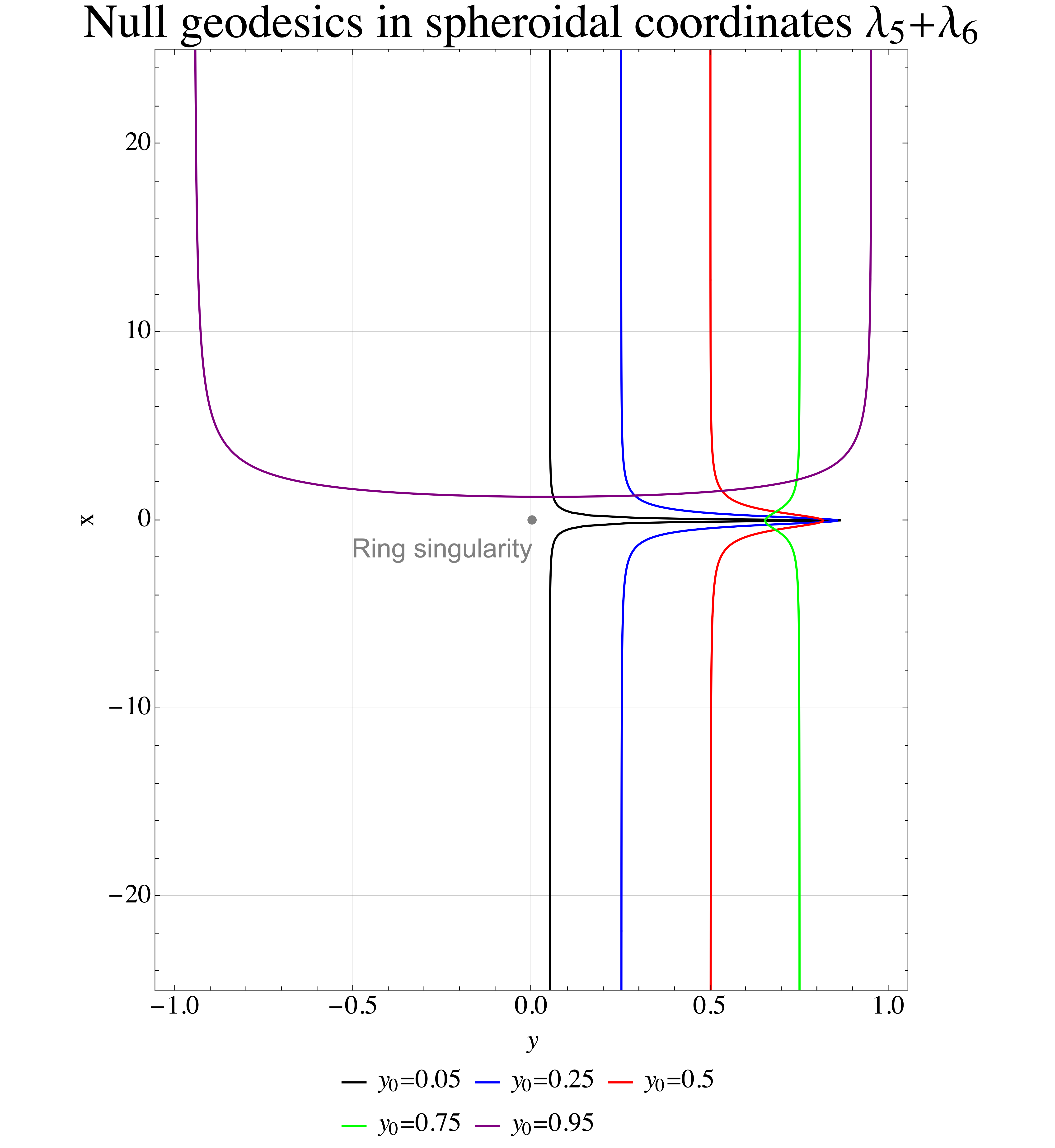}}
\subfloat[\centering]{\includegraphics[width=6cm]{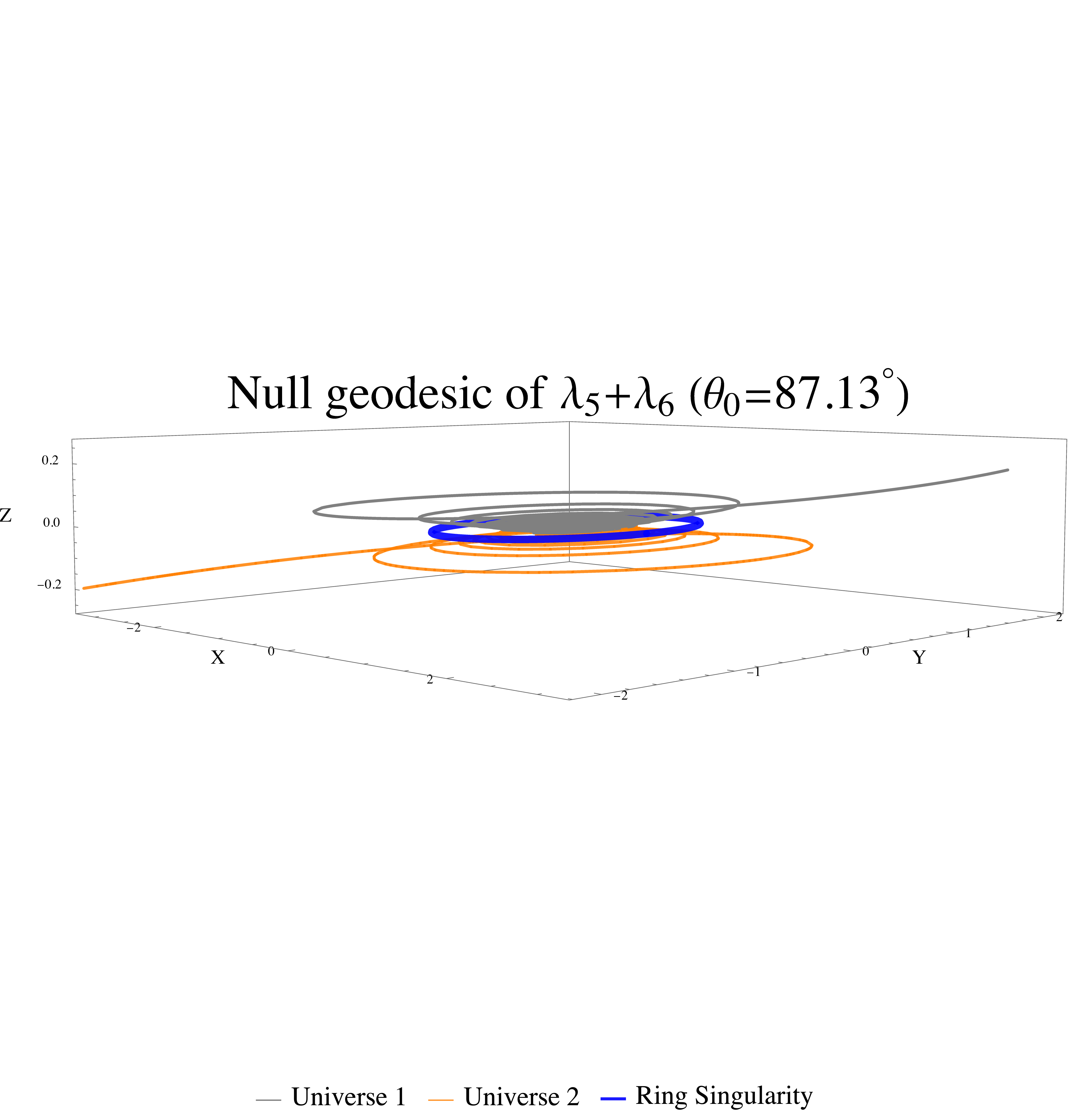}}\\
\subfloat[\centering]{\includegraphics[width=7.5cm]{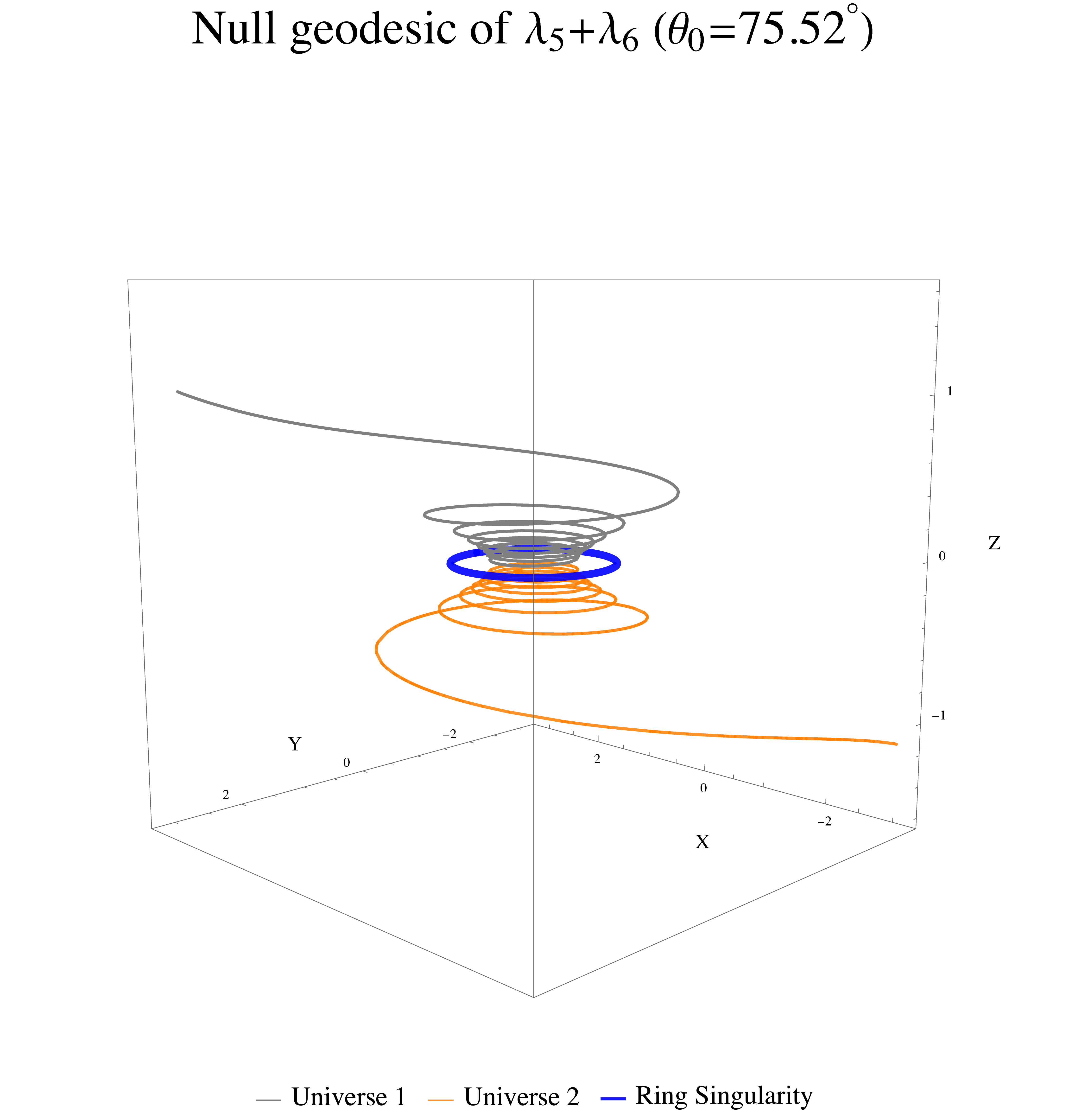}}
\subfloat[\centering]{\includegraphics[width=8.5cm]{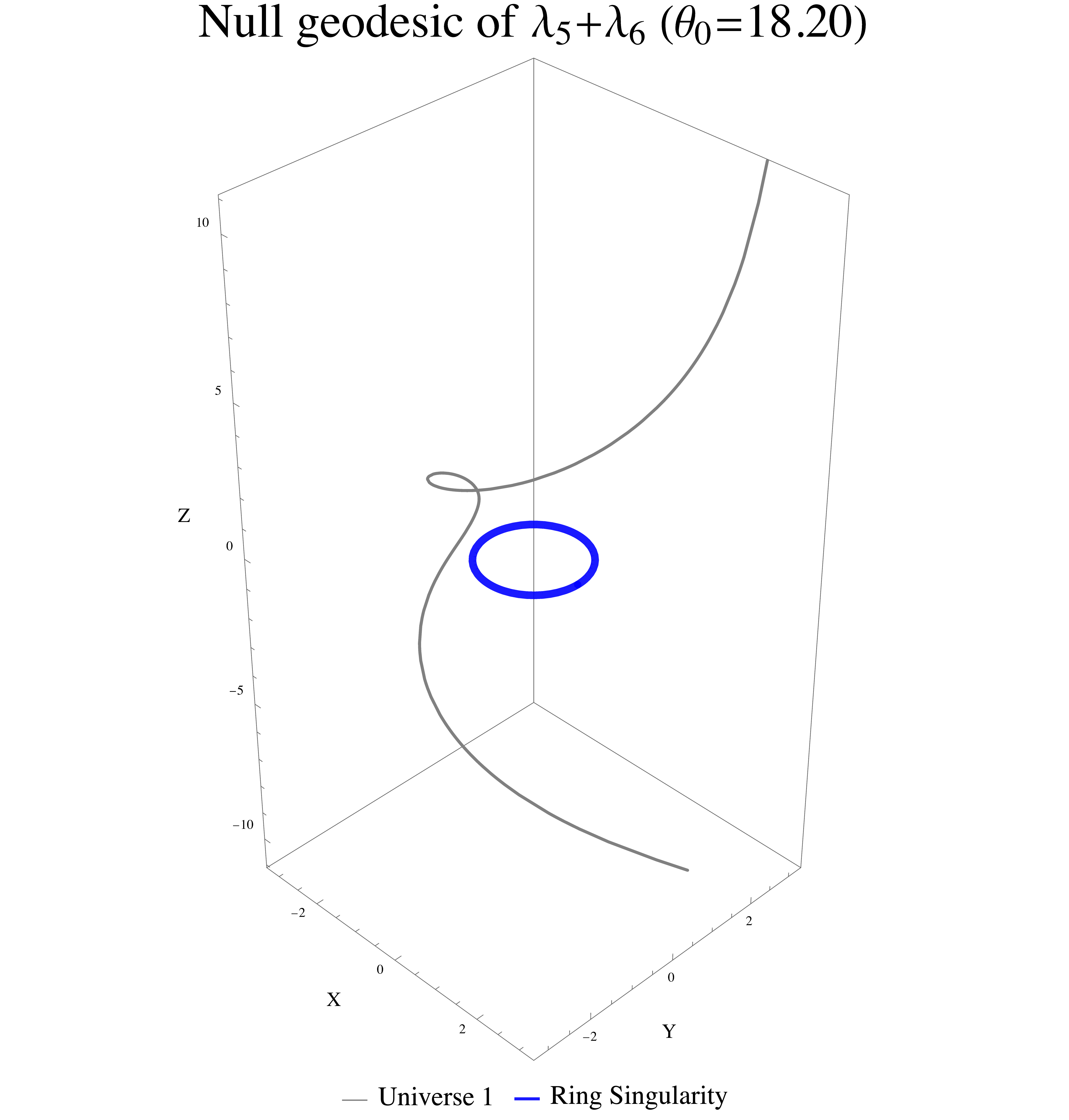}}

\caption{This 
 illustration is sourced from~\cite{Bixano:2025bio}. (a) This graph depicts null geodesics with varying initial values for $y$ using spheroidal coordinates, where $x>0$ represents one universe and $x<0$ signifies another. (b), (c), and (d) illustrate null geodesics in pseudo-Cartesian coordinates, the blue torus symbolizes the ring singularity, the orange colour denotes one universe, and the gray colour signifies another universe.}\label{fig:GeodesicasAll}
\end{figure}

Figure \ref{fig:GeodesicasAll}a illustrates the plotted null geodesics corresponding to various values of $y_0$ within spheroidal coordinates. Positive instances of $x$ are indicative of one universe, whereas negative instances of $x$ are associated either with a separate universe or with the same universe, but~in a different spatial~domain.

The remaining three figures in Figure~\ref{fig:GeodesicasAll} illustrate various null geodesics represented in pseudo-Cartesian coordinates as described in (\ref{pseudocartesian}). Universe 1 is denoted in gray, while Universe 2 is represented in orange. The~ring singularity is depicted as a blue torus. As~demonstrated in Figure~\ref{fig:GeodesicasAll}a,d, the~geodesic with an initial value of $y_0=0.95$ was unable to traverse the wormhole and only approached~it.

\subsection{Vectorial Electromagnetic~Field}
A crucial and intriguing topic of examination is the structure of the vectorial electromagnetic field. Utilizing the representation of the electromagnetic field in Weyl coordinates, as~provided by
\begin{subequations}\label{CampoElectromagneticoxy}
\begin{equation}\label{CampoMagnetico}
    \begin{bmatrix} B_z \\ -B_\rho \end{bmatrix}=\frac{-1/\sqrt{\sigma_0}}{L(x^2+y^2)}\frac{\sqrt{f_0}}{2\kappa_0}
    \begin{bmatrix}
        x \sqrt{(x^2+1)(1-y^2)} & -y \sqrt{(x^2+1)(1-y^2)} \\
        y(x^2+1) & x(1-y^2)
    \end{bmatrix}
    \begin{bmatrix} \partial_{x} \\\partial_{y} \end{bmatrix} \left( \frac{\omega}{L} e^{-\lambda_c} \right),
\end{equation}
\begin{equation}\label{CampoElectrico}
    \begin{bmatrix} E_\rho \\ E_z \end{bmatrix}=\frac{c/\sqrt{\sigma_0}}{L(x^2+y^2)}\frac{\sqrt{f_0}}{2\kappa_0}
    \begin{bmatrix}
        x \sqrt{(x^2+1)(1-y^2)} & -y \sqrt{(x^2+1)(1-y^2)} \\
        y(x^2+1) & x(1-y^2)
    \end{bmatrix}
    \begin{bmatrix} \partial_{x} \\\partial_{y} \end{bmatrix} \left(  e^{-\lambda_c} \right),
\end{equation}
\end{subequations}
we are able to visualize the vectorial field. However, in~this instance, it is necessary to employ pseudo-Cartesian coordinates once more. For~this purpose, the~electromagnetic field will be projected in each direction $(u,v,w)$ provided by
\begin{align*}
    H_u(u,v,w)&=H_\rho(u,v,w) \cos{(\frac{d \varphi}{dt} t)} ,\\
    H_v(u,v,w)&=H_\rho(u,v,w) \sin{(\frac{d \varphi}{dt} t)} ,\\
    H_w(u,v,w)&=H_z(u,v,w),
\end{align*}
where the Boyer--Lindquist coordinates are mapped to  these pseudo-Cartesian coordinates as follows:
\begin{equation*}
     Lx+l_1=r=\sqrt{u^2+v^2+w^2}, \quad y=\cos{\theta}=w/r,
\end{equation*}

The equatorial plane $y=0$ is in alignment with plane $w=0$, signifying the presence of a ring singularity at points $w=0$ and $l_1^2=u^2+v^2$. Furthermore, it is essential to recognize that the internal values of the sphere with a radius $x=0\Rightarrow r=l_1$, as~established by $r<l_1$, are prohibited, thereby indicating that this region does not form part of the spacetime~continuum.

\begin{figure}[H]

\subfloat[\centering]{\includegraphics[width=6.8cm]{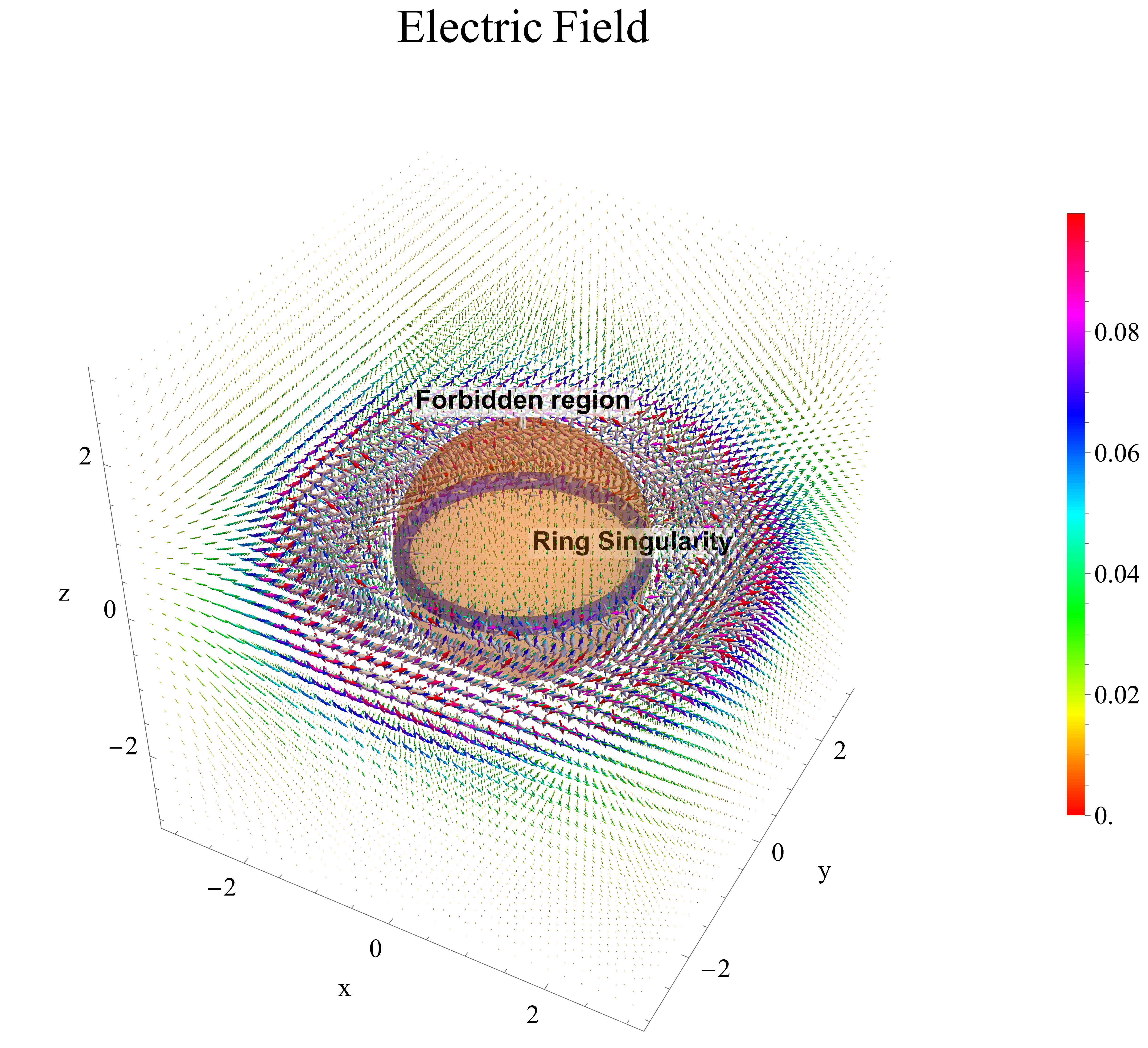}}
\subfloat[\centering]{\includegraphics[width=6.8cm]{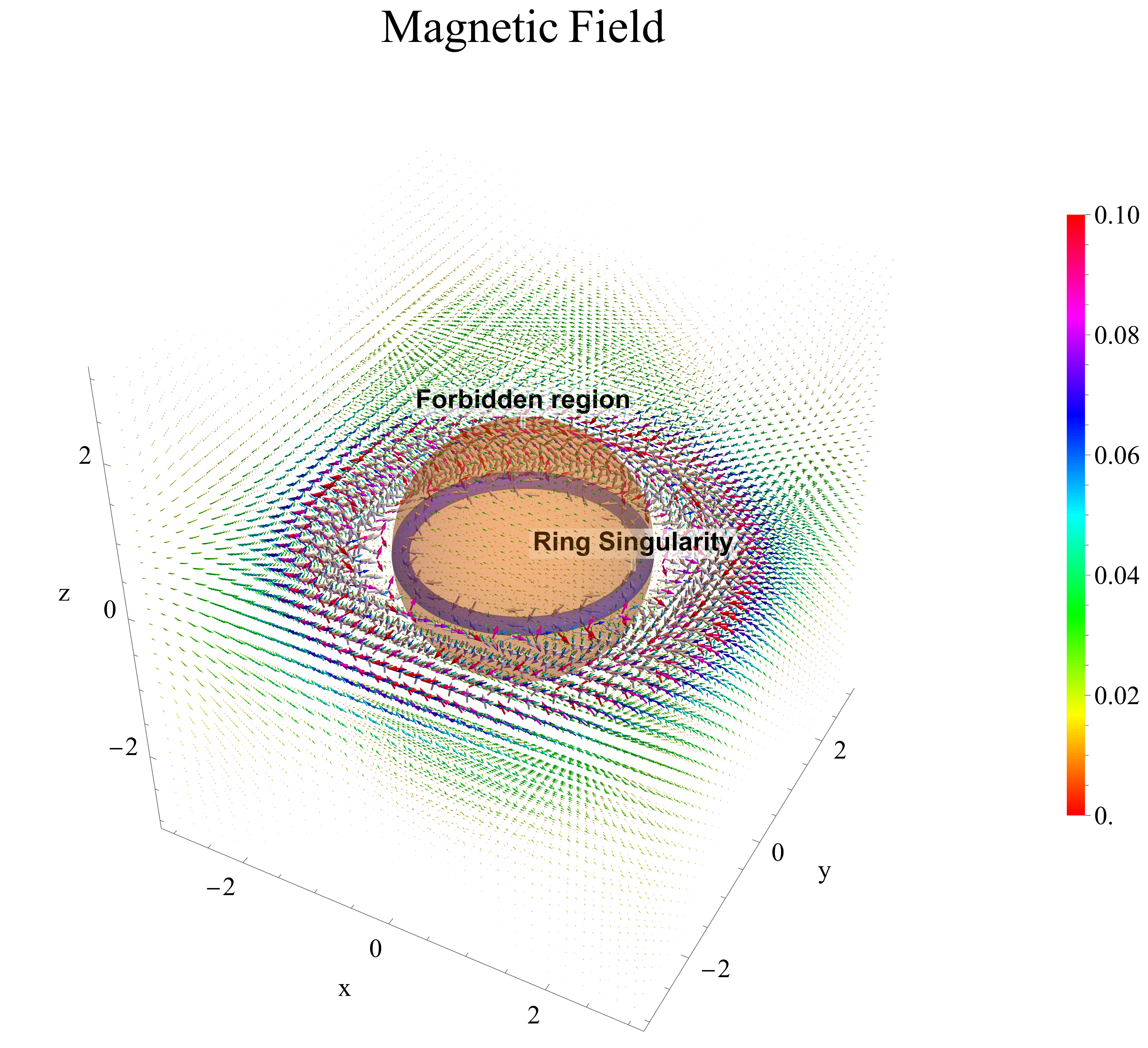}}

\caption{There illustrations are sourced from~\cite{Bixano:2025bio}. (a) depict the vectorial electric field using pseudo-Cartesian coordinates, and (b) illustrate the vectorial magnetic field. In both images, the ring singularity is represented by a blue torus, while the forbidden region is indicated by a red sphere. These graphs are purely schematic and feature a color intensity scale.}\label{fig:CampoElectromagnetico}
\end{figure}

Should one generate a plot of (\ref{CampoElectromagneticoxy}), the~outcome is represented by Figure \ref{fig:CampoElectromagnetico}. It is crucial to note that all graphs are schematic and not to scale, but~the key point is that while the electromagnetic field is remarkably strong near the ring singularity, where it diverges to infinity, the~navigation through the WH is still possible near to along the polar axis. The~singularity residing within the forbidden inner region corresponding to the sphere with radius $r=l_1$.

By focusing solely on the magnitude of the electromagnetic field, we have proposed a lower bound for the safe size of a wormhole. For~a wormhole with mass $M=10^{7}M_\odot $, the~maximum intensity of the magnetic field along the polar axis is approximately \mbox{$1700$~Teslas}, which is considered relatively safe.

\section{Causal~Structure}\label{sec4}
To establish the causal framework of this compact object, it is crucial to examine the smoothness and analyticity of (\ref{ds}). For~our purposes, it is preferable to employ spheroidal coordinates, as~we have identified the singularity and (\ref{SolucionLambdaCombinada}) is available in these~coordinates. 

In the second class of solutions, as~noted in (\ref{SegundaClaseSoluciones}), the~metric function $f$ remains constant, which implies a priori the absence of event horizons. Taking into account (\ref{OmegaxCero}) and (\ref{kxCero}) suggest that these expressions preserve regularity for all $y \neq 0$ with $x=0$. Consequently, the~metric remains regular at the throat ($x=0$) for all $y_0 \neq 0$ and is asymptotically~flat. 

By utilizing the existence of axial symmetry, the~causal structure can be examined using only a subspace ($t,x,y$) (for further details, see~\cite{Chrusciel:2020fql}), i.e.,~fixing $\varphi=\varphi_0$, and~$f_0=1$ in (\ref{SolucionLambdaCombinada}), we obtain
\begin{align}\label{ds sin varphi}
    ds^2=-f c^2dt^2 +f^{-1} e^{2k} (d\rho ^2 + dz^2) =f c^2dt^2 +\frac{e^{2k}}{f}L^2(x^2+y^2) \left( \frac{dx^2}{x^2+1} + \frac{dy^2}{1-y^2}\right),
\end{align}

Considering $y=y_0$ for simplicity, and~defining the total 
 variable as $fdl^2= e^{2k} (d\rho^2+dz^2)/f $, we can employ the radial null variables $u=t-l$, $v=t+l$. Consequently, the~metric (\ref{ds sin varphi}) is transformed to $ds^2=-fdudv$, where $f=f_0$, then a regular metric has been obtained, allowing for the process of compactification to~proceed.

Consider the compactification given by $u = \tan{U} = t - l$, and~$v = \tan{V} = t + l$. The~line element is subsequently expressed as $ds^2 = -f(\sec{U} \sec{V})^2 dU dV$. By~applying the conformal transformation $\Omega \equiv \cos{U} \cos{V}$ and defining the variables $V = T + R$ and $U = T - R$, we derive the final form of the metric
\begin{equation}
    d\overline{s}^2 = \Omega^{-2} ds^2=-f(dT^2-dR^2).
\end{equation}

The boundaries in $\Omega=0$ correspond to the infinite future/past of the space and \mbox{the time}
\begin{align*}
    &u \rightarrow \pm \infty \quad \text{or} \quad U =\pm \pi/2 \qquad \text{implies} \qquad T=\pm \pi/2+R, \\
    &v \rightarrow \pm \infty \quad \text{or} \quad V =\pm \pi/2 \qquad \text{implies} \qquad T=\pm \pi/2-R,
\end{align*}

Through meticulous analysis (the complete process and explanation are given in~\cite{Bixano:2025qxp}), one can derive the behaviour close to the singularity as follows:
\begin{align}\label{ComportamientoGarganta}
    R(y_0,x=0) &\equiv R_{G} \neq 0, \\
    \lim\limits_{y_0 \rightarrow 0} R(y_0,x=0) &=\lim\limits_{y_0 \rightarrow 0} R_G=0,
\end{align}
and $T\in [-\pi/2,\pi/2]$. In~other words, in~the equatorial plane, the~throat is closed ($R_G(0,0)=0$).

\begin{figure}[H]

\centering
\includegraphics[width=15cm]{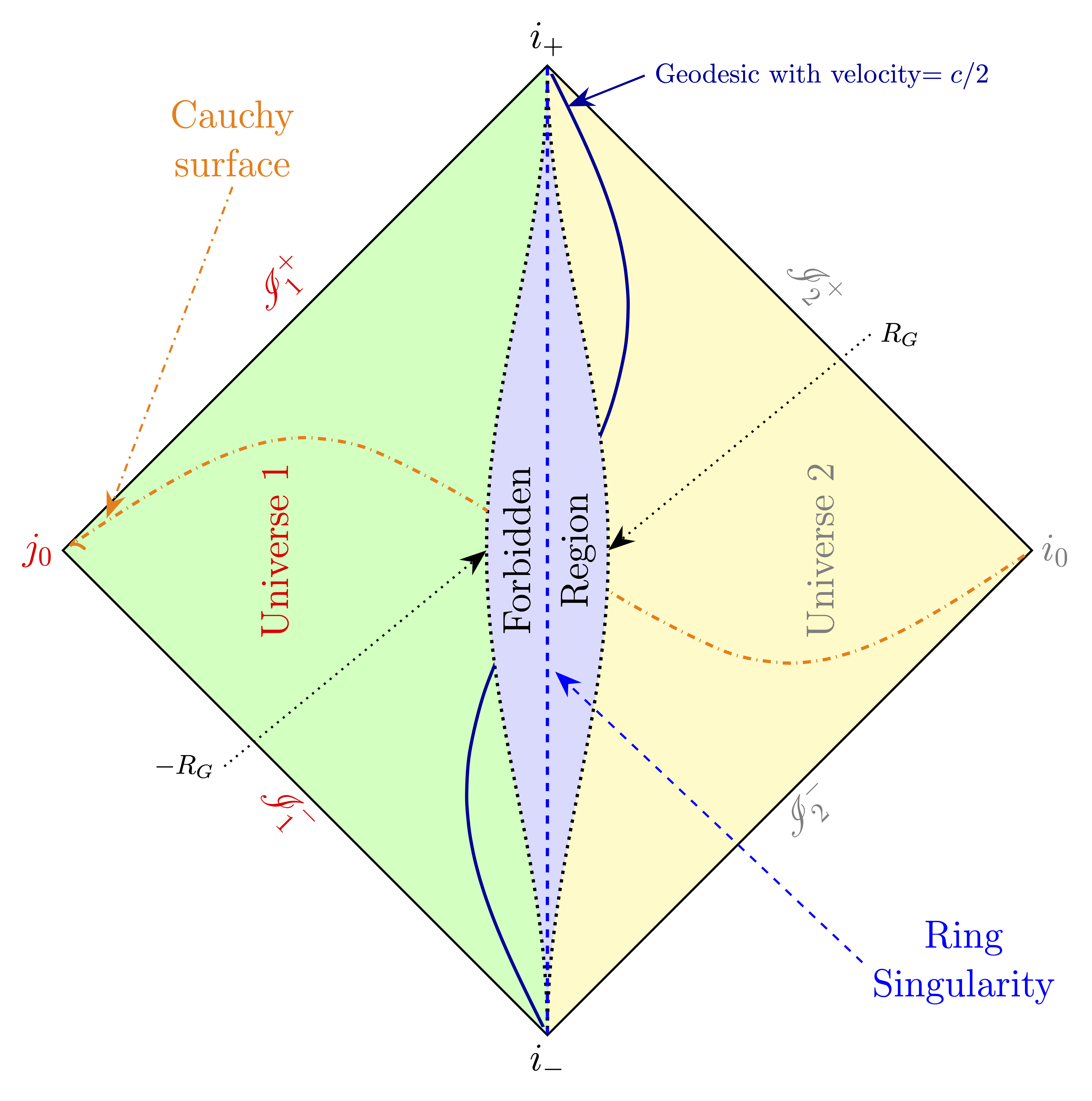}

\caption{This illustration is sourced from~\cite{Bixano:2025qxp}.The green area represents the Carter-Penrose diagram of one universe, while the yellow area pertains to another. The dotted black lines, which are topologically identified, form the throat in both universes. The dashed blue line depicts the ring singularity, and the orange dotted-dashed line represents a Cauchy surface.}\label{fig:EstructuraCausalDiagrama}
\end{figure}

Employing the constructed compactification, we have devised Figure \ref{fig:EstructuraCausalDiagrama}, which corresponds to the Carter--Penrose diagram for the WH (\ref{SolucionLambdaCombinada}). The~diagram comprises three distinct regions: the left-hand side depicted in green represents Universe 1, and~the region in yellow denotes Universe 2. Both regions have a future time-like infinity ($i_+$) and a past time-like infinity ($i_-$), with~their respective spatial infinities, $j_0$ and $i_0$, for~Universes 1 and 2. Additionally, past and future null infinities ($\mathscr{I}^{\pm}_{1,2}$) are present for both universes. The~third region, illustrated in blue, is the prohibited region, indicating values of $r<l_1$ and hiding the ring singularity. A~purple line delineates the geodesic of a particle with speed $c/2$, illustrating its traversal through the wormhole from one universe to the~other.

In Boyer--Linquist coordinates, the~throat is a $\mathbb{S}^2$ sphere with a radius of $r=l_1$ and the singularity is in ($r=l_1,\theta=\pi/2$). The~$\mathbb{S}^2$ sphere of radius $r=l_1$ is topologically connected to another $\mathbb{S}^2$ sphere of radius $r=-l_1 \quad \forall \theta$. In~Weyl-coordinates, the~throat is at $z=0$, and~the $z=0^{+}$ plane is topologically identified as $z=0^{-}$, in~this case, the~ring singularity is in ($\rho=L,z=0$). The~black dotted lines represent the throat ($\pm R_G$), which is topologically identified and corresponds to the aforementioned~descriptions.

\begin{quote}
    It is now understood that the throat \textit{dresses} the ring singularity similarly to how event horizons hide singularities in black holes. The~mechanics are as follows: when one attempts to approach the singularity in the equatorial plane, the~wormhole tends to constrict its throat. Consequently, the~compact objects prevent traversal to another universe. Moreover, according to $\theta \neq \phi/2$, this will never pose a problem, because~as we approach the ring singularity, we first encounter the throat, which consequently transports us into another universe.
\end{quote}

Using the definitions of Killing horizons, Cauchy horizons, and~event horizons as provided in reference~\cite{Chrusciel:2020fql}, we can examine the solution \eqref{SolucionLambdaCombinada}. 

We commence our analysis on the absence of event horizons. In~order to discern the event horizons within these solutions, we shall employ the Killing vector $K_t=\partial_t + \Omega\partial_\psi$, where $\Omega$ is constant over the horizon $\mathscr{H}$. This will be achieved by computing the norm of the said vector $K_t$:
\begin{align*}
    &g(K_t,K_t)=g_{tt}+2\Omega g_{t\psi}+\Omega^2g_{\psi \psi}=0, \\
    & \Rightarrow \Omega_{\pm}=-\frac{g_{t\psi}}{g_{\psi \psi}}\pm \frac{\sqrt{-(g_{\psi \psi}g_{tt}-(g_{t\psi})^2)}}{g_{\psi \psi}},
\end{align*}

Consequently, the~event horizon can be determined $\mathscr{H}$ provided that $\Omega$ are non-degenerate and remain~constant. 
\begin{subequations}
\begin{align}
    &g_{\psi \psi}g_{tt}-(g_{t\psi})^2 =-\rho=-L\sqrt{(x^2+1)(1-y^2)}=0\label{Horizonte condicion} ,\\
    &\Omega=\Omega_-=\Omega_+=-\frac{g_{t\psi}}{g_{\psi \psi}}=-(\omega|_{\mathscr{H}})^{-1}
\end{align}
\end{subequations}

The sole feasible event horizon is located in the polar axis $y=\pm1$, characterized by $\omega|_{\mathscr{H}}=-L\tau_0$. However, upon~opting for the asymptotically flat solution $\tau_0=0$, the~apparent event horizon~vanishes.

A null hyper-surface $G(x,y)$ is defined by the existence of a normal vector $\zeta$, which adheres to the condition $\zeta^2=0$ over the hyper-surface. Thus, a~Killing horizon is characterized as a null hypersurface that contains a Killing vector $\xi$, which is concurrently null $\xi^2=0$ and tangent $\xi^{\mu} \zeta_{\mu} = 0$ throughout $G(x,y)$.
Let
\begin{equation}
    G(x,y) \equiv \rho^2-\omega^2 = 0 \qquad \text{implies} \qquad \omega|_G = \pm \rho,
\end{equation}
then, the~normal vector~is
\begin{equation}
    \zeta \equiv (g^{xx}\partial_x G)\partial_x+(g^{yy}\partial_yG)\partial_y =e^{-2k}\big( \partial_x G \partial_x + \partial_y G \partial_y \big),
\end{equation}
where condition $\zeta^2|_G=0$ is satisfied.
Selecting the killing vector $\xi=\partial_\varphi$, and~computing
\begin{equation}
    \xi^2|_{G}=g(\partial_\varphi,\partial_\varphi)|_{G}=g_{\varphi \varphi}|_{G}=\rho^2-\omega^2|_G=0, \qquad \xi^{\mu} \zeta_{\mu} = 0.
\end{equation}

It is observable that $\xi \propto \zeta $, indicating that $\xi$ acts as both the normal and tangent vector to the hypersurface $\omega = \pm \rho$. It is crucial to clarify that this horizon signifies the boundary at which causality violation,\textbf{ it is not a event horizon}. 

The surface gravity is provided by $\nabla_\mu (\xi ^2)|_G =-2 \varrho \xi_{\mu} $, and~employing the one-form Killing vector $\xi^{\sharp}=g_{t \varphi}dt+g_{\varphi \varphi}d\varphi$, it is possible to determine
\begin{equation} 
\begin{split}
    d(\xi^2)&=d(g_{\varphi \varphi})=(\partial_x g_{\varphi \varphi})dx+(\partial_y g_{\varphi \varphi})dy, \\
    &=2(\omega\partial_x \omega-\rho \partial_x \rho)dx+2(\omega\partial_y \omega-\rho \partial_y \rho)dy.
\end{split}
\end{equation}
Therefore $d(\xi^2)|_{G}=0 \Rightarrow\varrho=0$, indicating zero surface gravity at this Killing horizon, allowing an observer at infinity to cross it~easily.

Now, taking anvantage of the norm of the Killing vector $\partial_\varphi$, we can see some important~things

\begin{itemize}
    \item The Killing vector $K_\varphi=\partial_\varphi$ is space-like if and only if $g(K_\varphi,K_\varphi)=g_{\varphi \varphi}=\rho^2-\omega^2 \geq 0$.
    \item $t$ is a temporal-type function if and only if $g(\nabla t,\nabla t)=g(g^{t\nu}\partial_\nu ,g^{t\mu}\partial_\mu)=g^{t\nu}g^{t\mu}g_{\mu \nu}=-(\rho^2-\omega^2)/\rho^2 < 0$, then $\rho^2-\omega^2>0$.
\end{itemize}

Based on the aforementioned evidence and the analysis provided in~\cite{Lobo:2002rp}, it can be ascertained that the \textit{Closed Timelike Curves} (CTC) manifest under the condition\linebreak $g_{\phi \phi}=\rho^2-\omega^2<0$ in the context of our specific~study.

\begin{quote}
    After conducting some analysis, it is evident that $\rho^2 > g_{\phi \phi} = \rho^2 - \omega^2 \quad \forall \{x, y\}$. Consequently, before~an entity can interact with points where CTCs manifest, the~entity traverses the wormhole into another universe. A~similar phenomenon occurs with the Kerr--Newman black hole, where a ring singularity exists, and~CTCs emerge nearby. However, the~event horizon conceals this occurrence, enabling us to apply the solution to the exterior regions, outside the event horizon. In~our scenario, the~throat regulates the solution, and~we consider the entire object because the throat is closed in the equatorial plane. Before~an object can approach the singularity or CTCs appear, it passes through the wormhole.
\end{quote}



\section{Flat~Subspaces method}
\label{section: flat subspaces}

In this section, we will briefly describe a method for solving the vacuum Einstein field equations (EFEs) in higher~dimensions.

In~\cite{Sarmiento-Alvarado2025}, the~authors consider a spacetime with a metric $\hat{g}$, which admits $n$ commutative Killing vectors.
Under this assumption, the~metric $\hat{g}$ can be written as
\begin{equation}
    \hat{g}
    = f ( \upsilon, \zeta ) ( d\upsilon^2 + d\zeta^2 )
    + g_{i j} ( \upsilon, \zeta ) dx^i dx^j
\end{equation}
for all $i, j \in \{ 3, \ldots, n + 2 \}$.
In a vacuum, the~Einstein field equations are equivalent to $R_{A B} = 0$ for all $A, B \in \{ 1, \ldots, n + 2 \}$.
Then,
\begin{align}
\label{chiral eq g}
&    ( \upsilon g_{, w} g ^{-1} )_{, \bar w} + ( \upsilon g_{, \bar w} g ^{-1} )_{, w} = 0
,
\\\label{SL invariant field eq f}
&
    ( \ln f \upsilon ^{1-1/n} )_{, W} = \frac{1}{2} \upsilon \operatorname{tr} ( g_{, _W} g^{-1} )^2
    \text{ for } W \in \{ w, \bar w \},
\end{align}
where $g$ is defined as $( g )_{i j} = -\upsilon^{-2/n} g_{i j}$, $\upsilon = \sqrt{ -\det g_{i j}}$ and $w = \upsilon + i \zeta$.
Note that $g$ is a symmetry matrix and belongs to the Lie group $SL ( n, \mathbb{R} )$.

In order to solve the chiral Equation \eqref{chiral eq g}, they assume that $g$ depends on a set of parameters $\{ \xi^a ( w, \bar{w} )\}$, which satisfy the generalized Laplace equation
\begin{equation}
\label{gen Laplace eq}
    ( \upsilon \xi^a_{, w} )_{, \bar{w}} + ( \upsilon \xi^a_{, \bar{w}} )_{, w} = 0 .
\end{equation}

Then, the~chiral Equation \eqref{chiral eq g} becomes
\begin{equation}
\label{Killing eq}
    A_{a, b} + A_{b, a} = 0 \text{ for all } a, b \in \{ 1, \ldots, r \} ,
\end{equation}
where $A_a = g_{, a} g^{-1}$ and $r \in \{ 1, \ldots, n \}$.
The partial derivative of $A_a$ with respect to $\xi^b$ is given by $A_{a, b} = -\frac{1}{2} [ A_a, A_b ]$.
They suppose that the matrices $A_a$ are constant.
Consequently, $\{ A_a \}$ is a set of pairwise commuting~matrices.

Since the determinant of $g$ is constant, the~matrices $A_a$ are traceless; hence, they belong to the Lie algebra $\mathfrak{sl} ( n, \mathbb{R} )$.
The matrices $A_a$ are transformed as $A_a \to \mathfrak{C} A_a \mathfrak{C}^{-1}$ when $g \to \mathfrak{C} g \mathfrak{C}^T$ with constant $\mathfrak{C} \in SL ( n, \mathbb{R} )$.
Thus, we can define an equivalence relation under this transformation and partition the set $\{ A_a \}$ into equivalence classes.
In~\cite{Sarmiento-Alvarado2023}, a~classification of the equivalence classes of $\mathfrak{sl} ( n, \mathbb{R} )$ is given, which is based on their types of~eigenvalues.

In~\cite{Sarmiento-Alvarado2025}, the~authors assume that $A_1$ is representative of some equivalence classes of $\mathfrak{sl} ( n, \mathbb{R} )$.
Then, $A_2$ is chosen from the centralizer of $A_1$, i.e.,~$\mathcal{C} ( A_1 )$, and~so on, until~$A_r$ is chosen from $\mathcal{C} (\{ A_1, \ldots, A_{r - 1} \})$.
Since computing the centralizer of large and complex matrices is difficult, the~same authors propose an alternative.
They demonstrate that there exists a commutative algebra $\mathfrak{A}$ of dimension $n$ contained in the centralizer of a representative of some equivalence classes of $\mathfrak{sl} ( n, \mathbb{R} )$.
In this way, $\{ A_a \}$ is a subset in $\mathfrak{A}$.
Hence, the~matrices $A_a$ are selected from $\mathfrak{A}$.
Furthermore, they compute the centralizers of the equivalence classes of $\mathfrak{sl} ( n, \mathbb{R} )$ and determine the commutative algebras contained in~them.

Let $\{ A_a \} \subset \mathfrak{sl} ( n, \mathbb{R} )$ be a set of pairwise commuting matrices, and~let $\{ \xi^a ( w, \bar{w} ) \}$ be a set of solutions to the generalized Laplace equation.
Then, the~solution $g ( w, \bar{w} )$ is given by
\begin{equation}
\label{chiral sol}
    g ( w, \bar{w} ) = \exp( \xi^a ( w, \bar{w} ) A_a ) g_0 ,
\end{equation}
where $g_0 \in \mathcal{I} (\{ A_a \}) = \{ M \in \mathbf{Sym}_n : A_a M = M A_a^T \text{ for all } a \in \{ 1, \ldots, r \} \}$.
Now, \mbox{Equation~\eqref{SL invariant field eq f}} can expressed in terms of the matrices $A_a$ and the parameters $\xi^a ( w, \bar{w} )$ as
\begin{equation}
    ( \ln f \upsilon^{1 - 1/n} )_{, W}
    = \frac{\operatorname{tr} A_a A_b }{2} \upsilon \xi^a_{, W} \xi^b_{, W} .
\end{equation}
The solutions of chiral equations given by $g ( w, \bar{w} ) = \exp( \xi ( w, \bar{w} ) A ) g_0$ can be found in~\cite{Sarmiento-Alvarado2023}.

\section{A Family of Wormholes in~5D}   %
\label{section: family wormholes}

In this section, we will build an exact solution to the 5-dimensional EFE using the method introduced in the previous section and the mathematical tools from~\cite{Sarmiento-Alvarado2025}.

Let
\begin{equation}
    A_1 = \left[\begin{array}{lll}
        1 && \\
        & 1 & \\
        && -2 \\
    \end{array}\right] \text{ and }
    A_2 = \left[\begin{array}{lll}
        & -1 &\\
        1 &&\\
        && 0 \\
    \end{array}\right]
\end{equation}
be a pair of pairwise commuting matrices in $\mathfrak{sl} ( 3, \mathbb{R} )$.
Let
\begin{equation}
    \xi^1
    = \xi^1_0
    -\frac{2}{3} \ln \upsilon
    + q \tan^{-1} \frac{r}{r_0}
    \text{ and }
    \xi^2
    = \xi^2_0
    + p \tan^{-1} \frac{r}{r_0}
\end{equation}
be a pair of solutions to the generalized Laplace equation
\begin{equation}
\label{Laplace eq Boyer-Lindquist coordinates}
    ( ( r^2 + r_0^2 ) \xi^a_{, r} ) _{, r} + \frac{1}{\sin \theta} ( \xi^a_{, \theta} \sin \theta ) _{, \theta} = 0 ,
\end{equation}
which is written in terms of the Boyer--Lindquist coordinates as follows:
\begin{equation}
\label{def Boyer-Lindquist coordinates}
    \upsilon = \sqrt{ r^2 + r_0^2 } \sin \theta
    \text{ and }
    \zeta = r \cos \theta ,
\end{equation}
where $r \geq 0$,
$0 \leq \theta \leq \pi$,
$r_0$ is a positive number,
$p$ is a natural number, and~$q, \xi^1_0, \xi^2_0$ are real numbers.
Since $g_0$ is a constant matrix in $\mathcal{I} (\{ A_1, A_2 \})$, then it has the form
\begin{equation}
    g_0 = -\operatorname{diag} \left[
        \left[\begin{array}{rr}
            C_t & D \\
            D & -C_t \\ 
        \end{array}\right],
        C_\phi
    \right],
\end{equation}
where $C_t, D, C_\phi \in \mathbb{R}$ are constants.
To build a solution to the chiral equation, we compute Equation~\eqref{chiral sol}.
Therefore,
\begin{equation}
    g ( r, \theta )
    = -\rho^{-\frac{2}{3}} \operatorname{diag} \left[
        \Xi \left[\begin{array}{rr}
             \mathscr{U}_p ( \frac{r}{r_0} ) & \mathscr{V}_p ( \frac{r}{r_0} ) \\
             \mathscr{V}_p ( \frac{r}{r_0} ) & -\mathscr{U}_p ( \frac{r}{r_0} )
        \end{array}\right] ,
        \frac{ \rho^2 \Xi^{-2} }{\mu^2 + \nu^2}
    \right] ,
\end{equation}
where
\begin{align}
    \Xi
&   = \frac{
        \exp\left( q \tan^{-1} \frac{r}{r_0} \right)
    }{
        \exp\left( q \frac{\pi}{2} \right)
    } ,
\\  \mathscr{U}_p ( x )
&   = \mu T_p \left( \frac{1}{\sqrt{ x^2 + 1 }} \right)
    - \frac{\nu x}{\sqrt{ x^2 + 1 }} U_{p - 1} \left( \frac{1}{\sqrt{ x^2 + 1 }} \right),
\\  \mathscr{V}_p ( x )
&   = \nu T_p \left( \frac{1}{\sqrt{ x^2 + 1 }} \right)
    + \frac{\mu x}{\sqrt{ x^2 + 1 }} U_{p - 1} \left( \frac{1}{\sqrt{ x^2 + 1 }} \right)
\end{align}
for all $x \in \mathbb{R}$, $\mu = C_t \cos \xi^2_0 - D \sin \xi^2_0$, $\nu = D \cos \xi^2_0 + C_t \sin \xi^2_0$, $T_p$ and $U_{p - 1}$ are Chebyshev polynomials of the first and second kind, respectively.

In what follows, we present some properties of the functions $\mathscr{U}_p$ and $\mathscr{V}_p$.
\begin{itemize}

\item Recurrence relations.
\begin{equation}
\left\{
\begin{aligned}
\mathscr{U}_p(x) &= \frac{2}{\sqrt{x^2+1}}\,\mathscr{U}_{p-1}(x) - \mathscr{U}_{p-2}(x),\\
\mathscr{V}_p(x) &= \frac{2}{\sqrt{x^2+1}}\,\mathscr{V}_{p-1}(x) - \mathscr{V}_{p-2}(x),
\end{aligned}
\right.
\qquad (p>2).
\end{equation}

\item Pythagorean relation.
\begin{equation}
    \mathscr{U}_p^2 ( x )
    + \mathscr{V}_p^2 ( x )
    = \mu^2
    + \nu^2 .
\end{equation}

\item Roots.
\begin{equation}
\begin{aligned}
x_k &= \tan\!\left(\frac{\tan^{-1}(\mu/\nu)+k\pi}{p}\right)
&& \text{for } \mathscr{U}_p,\\
x_k &= \tan\!\left(\frac{-\tan^{-1}(\nu/\mu)+k\pi}{p}\right)
&& \text{for } \mathscr{V}_p,
\end{aligned}
\qquad \text{where } k\in\{0,\ldots,p-1\}.
\end{equation}

\item Limits.
\begin{equation}
\label{limits UV}
\begin{array}{lll}
    \lim _{x \to \pm \infty} \mathscr{U}_p ( x )
    = (-)^\frac{p}{2} \mu ,
&   \lim _{x \to \pm \infty} \mathscr{V}_p ( x )
    = (-)^\frac{p}{2} \nu
&   \text{for $p$ even},
\\  \lim _{x \to \pm \infty} \mathscr{U}_p ( x )
    = \mp (-)^\frac{p - 1}{2} \nu ,
&   \lim _{x \to \pm \infty} \mathscr{V}_p ( x )
    = \pm (-)^\frac{p - 1}{2} \mu
&   \text{for $p$ odd.} \\
\end{array}
\end{equation}

\item Differentiation.
\begin{equation}
\begin{array}{ll}
    \frac{d \mathscr{U}_p}{d x} ( x )
    = \frac{-p}{ x^2 + 1 } \mathscr{V}_p ( x ) ,
&   \frac{d \mathscr{V}_p}{d x} ( x )
    = \frac{p}{ x^2 + 1 } \mathscr{U}_p ( x ) .
\end{array}
\end{equation}

\end{itemize}

Solving the partial differential equations
\begin{equation}\label{diff eq f Boyer-Lindquist coordinates}
\begin{aligned}
    \left( \ln f \upsilon^{1-1/n} \right)_{, r} & = \frac{\operatorname{tr} A_a A_b}{4} \frac{
        r \sin ^2 \theta
    }{ \Delta } \Bigg[
        \xi^a_{, r} \xi^b_{, r} 
        + \frac{
            2 \cot \theta \xi^a_{, r} \xi^b_{, \theta}
        }{r}
        - \frac{
            \xi^a_{, \theta} \xi^b _{, \theta}
        }{ r^2 + r_0^2 }
    \Bigg] ,
\\  \left( \ln f \upsilon^{1-1/n} \right)_{, \theta}
&   = - \frac{\operatorname{tr} A_a A_b}{4} \frac{
        \sin \theta \cos \theta
    }{ \Delta } \Bigg[
        \xi^a_{, r} \xi^b_{, r}
   - \frac{
            2 r \tan \theta \xi^a_{, r} \xi^b_{, \theta}
        }{
            r^2 + r_0^2
        }
        - \frac{
            \xi^a_{, \theta} \xi^b_{, \theta}
        }{
            r^2 + r_0^2
        }
    \Bigg] ,
\end{aligned}
\end{equation}
we find
\begin{equation}
    f  ( r, \theta )
    = \Delta^\frac{3 q^2 - p^2}{4}
    \Xi^{-2} ,
\end{equation}
where
\begin{equation}
    \Delta = \frac{
        r^2 + r_0^2 \cos^2 \theta
    }{ r^2 + r_0^2 } . 
\end{equation}
Therefore, an~exact solution is
\begin{equation}
\label{wormhole 5D}
\begin{aligned}
    \hat{g}_p =
&   - \mathscr{U}_p (r/r_0) \Xi dt^2
    + \frac{
        \Delta^l (
        dr^2
        + ( r^2 + r_0^2 ) d\theta^2
    )
    + ( r^2 + r_0^2 ) \sin^2 \theta d\phi^2
    }{\Xi^2} 
\\& + \mathscr{U}_p (r/r_0) \Xi d\psi^2
    + 2 \mathscr{V}_p (r/r_0) \Xi dt d\psi
\end{aligned}
\end{equation}
where $l = 1 + \frac{3}{4} q^2 - \frac{1}{4} p^2$ and $\psi$ is the coordinate of the fifth dimension.
In the limit $r \to \infty$, if $\mathscr{U}_p (r/r_0) \to 1$ and $\mathscr{V}_p (r/r_0) \to 0$, then $\hat{g}_p \to - dt^2 + dr^2 + r^2 d\Omega_2^2 + d\psi^2$, where $d\Omega_2^2 = d\theta^2 + \sin^2 \theta d\phi^2$.

For the metric \eqref{wormhole 5D} to describe a wormhole, we extend the values of $r$; that is, $r \in \mathbb{R}$. 
However, we reject the odd values of $p$, because~$\mathscr{U}_p (r/r_0) \to -1$ as $r \to -\infty$.
This means that $t$ and $\psi$ are spatial and temporal coordinates, respectively, at~infinity.
When $\mu = -1$, $\nu = 0$ and $q = 0$, the metric $\hat{g}_2$ is
\begin{equation}
\label{wormhole g2}
    \hat{g}_2 =
    \frac{r^2 - r_0^2}{r^2 + r_0^2} ( d\psi^2 - dt^2 )
    + dr^2
    + ( r^2 + r_0^2 ) d\Omega_2^2
    - \frac{4 r_0 r}{r^2 + r_0^2} dt d\psi ,
\end{equation}
which describes a wormhole~\cite{Dzhunushaliev:1998rz, LU2008511}.
For this metric, the~Kretschmann invariant is
\begin{equation}
    R_{A B C D} R^{A B C D} = 24 \frac{( r_0^2 -2 r^2 ) r_0^2}{( r^2 + r_0^2 )^{4}} .
\end{equation}


\section{Conclusions}

The second family of solutions function as wormholes, and~it is crucial to note that the throat of the wormhole hides all the potential irregularities that may exist in the flat solution. In~this compact object, there is an absence of event horizons and ergo-regions, yet closed time-like curves (CTCs) do appear within the blue region of Figure~\ref{fig:EstructuraCausalDiagrama}. Essentially, one traverses the wormhole prior to encountering the singularity or the causal violation region. The~wormhole described herein adheres to the null energy condition (NEC) in the context of a dilaton scalar field and possesses a ring singularity without necessitating asymptotic flatness. We suggest specific bounds to guarantee the stability and safety of the~wormhole.

In Section~\ref{section: flat subspaces}, we briefly explain the flat subspaces method, which allows us to build exact solutions to the vacuum EFE in higher dimensions.
Its main advantage over other methods lies in the use of algebraic techniques.
In this approach, an~exact solution to the EFE is derived from the solution to the chiral equation.
Specifically, it only requires a set $\{ A_a \}$ of pairwise commuting matrices, a~set $\{ \xi^a ( w, \bar{w} ) \}$ of solutions to the generalized Laplace equation and a constant matrix $g_0 \in \mathcal{I} (\{ A_a \})$ to compute a solution of the chiral equation.
In Section~\ref{section: family wormholes}, we built a family of wormholes in five dimensions.
However, only metric $\hat{g}_2$, given by Equation~\eqref{wormhole g2}, was studied in detail.
Currently, we study the other values of $p$.

\vspace{12pt}

All authors have contributed equally to the conception, analysis, and preparation of this work.

This work was also partially supported by SECIHTI M\'exico under grants  A1-S-8742, 304001, 376127, 240512.

The authors declare that there are no conflicts of interest.

\appendix
\section[\appendixname~\thesection]{}
\subsection*{Parameter} 
 Constraint~Equation

For obtain the parameter constrain we will use the~metric 
\begingroup
\makeatletter\def\f@size{9}\check@mathfonts
\def\maketag@@@#1{\hbox{\m@th\normalsize\normalfont#1}}%
\begin{equation*}
ds^2= -f\left( d(ct)-\omega d \varphi \right)^2  + f^{-1} \bigg( L^2(x^2+1)(1-y^2) d\varphi^2 +L^2(x^2+y^2) e^{2k} \left\{ \frac{dx^2}{x^2+1} +\frac{dy^2}{1-y^2} \right\} \bigg),
\end{equation*}
\endgroup
where the function metrics associated to the solution \eqref{Lambda5+Lambda6} are
\begin{subequations}
\begin{align*}
        f&=f_0=1, \\
        \omega &=\frac{ L }{f_0} \bigg( \frac{\lambda_0 x(1-y^{2})-\tau_0 y (x^2+1)}{x^{2}+y^{2}} \bigg), \\
        A_\varphi &=\frac{\sqrt{f_0}}{2 \kappa_0 \sqrt{\sigma_0}} \bigg( A_{3}-\frac{\omega}{L}e^{-\lambda_{c}} \bigg),  \\
        A_t&=\frac{\sqrt{f_0}}{2\kappa_0 \sqrt{\sigma_0}} \bigg( e^{-\lambda_{c}} -1\bigg),\\
        k_{c} &= k_{\lambda5}+k_{\lambda6} -k_{0}\frac{8xy(1-y^2)(x^2+1)(x^2-y^2)\lambda_0 \tau_0}{4(x^2+y^2)^4} ,
\end{align*}
\end{subequations}
and the 4-potential $A_{\mu}=\bigg[ A_t(\rho,z),0,0,A_\varphi (\rho,z) \bigg]$.

By applying the variable transformation (See the articles~\cite{Matos:2000ai,Matos:2000za,Matos:2009rp})  $\kappa^2=e^{-2\alpha_0 \, \phi}$ and considering the scalar field potential described in Equation \eqref{SegundaClaseSoluciones}, we find~that
\[
\phi=-\frac{\ln{\kappa} }{\alpha_0}=-\frac{\lambda}{\alpha_0}\qquad | \qquad \kappa_0=1.
\]

Next, by~applying the Einstein Field Equations presented as~follows
\[
[EE]_{\mu \nu }\equiv R_{\mu \nu}-2\epsilon_0 \nabla_\mu \phi \nabla_\nu \phi  - 2 \sigma_0 e^{-2\alpha_0 \phi} \left( F_{\mu \sigma} \tensor{F}{_\nu}{^\sigma} -\frac{1}{4} g_{\mu \nu } F^2 \right),
\]
with $F_{\mu \nu }=\partial_{\mu}A_\nu-\partial_{\nu}A_\mu$, and~$F^2=F^{\mu \nu}F_{\mu \nu}$, we can replace the previously mentioned solution related to $\lambda_c$, resulting in
\begin{equation}
    [EE]_{\mu \nu }=\left(
\begin{array}{cccc}
 0 & 0 & 0 & 0 \\
 0 & [EE]_{xx } & [EE]_{xy } & 0 \\
 0 & [EE]_{yx } & [EE]_{yy} & 0 \\
 0 & 0 & 0 & 0 \\
\end{array}
\right)=0,
\end{equation}
where
\begin{subequations}\label{EE parameter constrain equation}
\begin{equation}
        [EE]_{xx }=\frac{\left(\alpha_0^2 (4 k_0+1)-4 \epsilon_0\right) \left(2 \lambda_0 x y+\tau_0 \left(x^2-y^2\right)\right)^2}{2 \alpha_0^2 \left(x^2+y^2\right)^4}=0,
    \end{equation}
\begin{equation}
        [EE]_{yy }=\frac{\left(\alpha_0^2 (4 k_0+1)-4 \epsilon_0\right) \left(\lambda_0 \left(y^2-x^2\right)+2 \tau_0 x y\right)^2}{2 \alpha_0^2 \left(x^2+y^2\right)^4}=0,
    \end{equation}
\begin{align}
        [EE]_{xy }=[EE]_{yx }&=-\frac{\left(\alpha_0^2 (4 k_0+1)-4 \epsilon_0\right)}{2 \alpha_0^2 \left(x^2+y^2\right)^4}\Bigg(2 \lambda_0^2 x y \left(x^2-y^2\right)\notag
        \\
        &+\lambda_0 \tau_0 \left(-6 x^2 y^2+x^4+y^4\right)+2 \tau_0^2 x y \left(y^2-x^2\right)\Bigg)=0,
    \end{align}
\end{subequations}

Therefore, \eqref{EE parameter constrain equation} is fulfilled if and only~if

\[
\left(\alpha_0^2 (4 k_0+1)-4 \epsilon_0\right)=0.
\]

The \textit{Maxwell Equation} 
 \eqref{Eq:Campo1} and the \textit{Klein--Gordon Equation} \eqref{Eq:Campo2} are satisfied regardless of the constraints on this~parameter.


\bibliographystyle{elsarticle-harv} 
\bibliography{Bibliografia}

@article{DelAguila:2018gni,
    author = "Del \'Aguila, Juan Carlos and Matos, Tonatiuh",
    title = "{Wormhole Cosmic Censorship: An Analytical Proof}",
    eprint = "1806.03747",
    archivePrefix = "arXiv",
    primaryClass = "gr-qc",
    doi = "10.1088/1361-6382/aaf336",
    journal = "Class. Quant. Grav.",
    volume = "36",
    number = "1",
    pages = "015018",
    year = "2019"
}

@article{Matos:2012gj,
    author = "Matos, Tonatiuh and Urena-Lopez, L. Arturo and Miranda, Galaxia",
    title = "{Wormhole Cosmic Censorship}",
    eprint = "1203.4801",
    archivePrefix = "arXiv",
    primaryClass = "gr-qc",
    doi = "10.1007/s10714-016-2040-7",
    journal = "Gen. Rel. Grav.",
    volume = "48",
    number = "5",
    pages = "61",
    year = "2016"
}

@article{Matos:2010pcd,
    author = "Matos, Tonatiuh",
    title = "{Class of Einstein-Maxwell Phantom Fields: Rotating and Magnetised Wormholes}",
    eprint = "0902.4439",
    archivePrefix = "arXiv",
    primaryClass = "gr-qc",
    reportNumber = "CIEA-09-GR5",
    doi = "10.1007/s10714-010-0976-6",
    journal = "Gen. Rel. Grav.",
    volume = "42",
    pages = "1969--1990",
    year = "2010"
}

@article{Matos:2000ai,
    author = "Matos, Tonatiuh and Nunez, Dario and Estevez, Gabino and Rios, Maribel",
    title = "{Rotating 5-D Kaluza-Klein space-times from invariant transformations}",
    eprint = "gr-qc/0001039",
    archivePrefix = "arXiv",
    reportNumber = "CINVESTAV-00-FIS-8",
    doi = "10.1023/A:1001982001694",
    journal = "Gen. Rel. Grav.",
    volume = "32",
    pages = "1499--1525",
    year = "2000"
}

@article{Matos:2000za,
    author = "Matos, Tonatiuh and Nunez, Dario and Rios, Maribel",
    title = "{Class of Einstein-Maxwell dilatons, an ansatz for new families of rotating solutions}",
    eprint = "gr-qc/0008068",
    archivePrefix = "arXiv",
    doi = "10.1088/0264-9381/17/18/323",
    journal = "Class. Quant. Grav.",
    volume = "17",
    pages = "3917--3934",
    year = "2000"
}

@article{Morris:1988cz,
    author = "Morris, M. S. and Thorne, K. S.",
    title = "{Wormholes in space-time and their use for interstellar travel: A tool for teaching general relativity}",
    doi = "10.1119/1.15620",
    journal = "Am. J. Phys.",
    volume = "56",
    pages = "395--412",
    year = "1988"
}

@article{Matos:2009rp,
    author = "Matos, Tonatiuh and Miranda, Galaxia and Sanchez-Sanchez, Ruben and Wiederhold, Petra",
    title = "{Class of Einstein-Maxwell-Dilaton-Axion Space-Times}",
    eprint = "0905.4097",
    archivePrefix = "arXiv",
    primaryClass = "gr-qc",
    reportNumber = "CIEA-09-GR14",
    doi = "10.1103/PhysRevD.79.124016",
    journal = "Phys. Rev. D",
    volume = "79",
    pages = "124016",
    year = "2009"
}

@article{DelAguila:2015isj,
    author = "Del \'Aguila, Juan Carlos and Matos, Tonatiuh and Miranda, Galaxia",
    title = "{Exact Rotating Magnetic Traversable Wormholes satisfying the Energy Conditions}",
    eprint = "1507.02348",
    archivePrefix = "arXiv",
    primaryClass = "gr-qc",
    doi = "10.1103/PhysRevD.99.124045",
    journal = "Phys. Rev. D",
    volume = "99",
    number = "12",
    pages = "124045",
    year = "2019"
}

@book{Cita:LorentzianWormholes,
   author = {Matt Visser},
   isbn = {978-1-56396-653-8},
   month = {9},
   publisher = {American Institute of Physics Melville, NY},
   title = {Lorentzian Wormholes: From Einstein to Hawking},
   year = {1996},
}

@article{DelAguila:2023twe,
    author = "Del \'Aguila, Juan Carlos and Matos, Tonatiuh",
    title = "{Geodesic completeness of a ring wormhole}",
    eprint = "2303.12251",
    archivePrefix = "arXiv",
    primaryClass = "gr-qc",
    doi = "10.1103/PhysRevD.107.064047",
    journal = "Phys. Rev. D",
    volume = "107",
    number = "6",
    pages = "064047",
    year = "2023"
}

@book{Lobo:2017cay,
    editor = "Lobo, Francisco S. N.",
    title = "{Wormholes, Warp Drives and Energy Conditions}",
    eprint = "2103.05610",
    archivePrefix = "arXiv",
    primaryClass = "gr-qc",
    doi = "10.1007/978-3-319-55182-1",
    isbn = "978-3-319-55181-4, 978-3-319-85588-2, 978-3-319-55182-1",
    publisher = "Springer",
    series = "Fundamental Theories of Physics",
    volume = "189",
    year = "2017"
}

@article{Bixano:2025jwm,
    author = "Bixano, Leonel and Matos, Tonatiuh",
    title = "{Einstein-Maxwell-dilaton wormholes that meet the energy conditions}",
    eprint = "2502.07206",
    archivePrefix = "arXiv",
    primaryClass = "gr-qc",
    doi = "10.1103/PhysRevD.111.084056",
    journal = "Phys. Rev. D",
    volume = "111",
    number = "8",
    pages = "084056",
    year = "2025"
}

@article{Minazzoli:2025nbi,
    author = "Minazzoli, Olivier and Wavasseur, Maxime",
    title = "{Compact objects with scalar charge embedded in a magnetic or electric field in Einstein\textendash{}Maxwell-dilaton theories}",
    eprint = "2502.13829",
    archivePrefix = "arXiv",
    primaryClass = "gr-qc",
    doi = "10.1140/epjc/s10052-025-14179-w",
    journal = "Eur. Phys. J. C",
    volume = "85",
    number = "4",
    pages = "474",
    year = "2025"
}

@book{Chrusciel:2020fql,
    author = "Chrusciel, Piotr",
    title = "{Geometry of Black Holes}",
    isbn = "978-0-19-887320-4, 978-0-19-885541-5",
    publisher = "Oxford University Press",
    series = "International Series of Monographs on Physics",
    month = "4",
    year = "2023"
}

@article{Penrose:1964wq,
    author = "Penrose, Roger",
    title = "{Gravitational collapse and space-time singularities}",
    doi = "10.1103/PhysRevLett.14.57",
    journal = "Phys. Rev. Lett.",
    volume = "14",
    pages = "57--59",
    year = "1965"
}

@article{Penrose:1969pc,
    author = "Penrose, R.",
    title = "{Gravitational collapse: The role of general relativity}",
    doi = "10.1023/A:1016578408204",
    journal = "Riv. Nuovo Cim.",
    volume = "1",
    pages = "252--276",
    year = "1969"
}

@book{Hawking:1973uf,
    author = "Hawking, Stephen W. and Ellis, George F. R.",
    title = "{The Large Scale Structure of Space-Time}",
    doi = "10.1017/9781009253161",
    isbn = "978-1-009-25316-1, 978-1-009-25315-4, 978-0-521-20016-5, 978-0-521-09906-6, 978-0-511-82630-6, 978-0-521-09906-6",
    publisher = "Cambridge University Press",
    series = "Cambridge Monographs on Mathematical Physics",
    month = "2",
    year = "2023"
}

@article{Hawking:1966sx,
    author = "Hawking, Stephen",
    title = "{The Occurrence of singularities in cosmology}",
    doi = "10.1098/rspa.1966.0221",
    journal = "Proc. Roy. Soc. Lond. A",
    volume = "294",
    pages = "511--521",
    year = "1966"
}

@article{Hawking:1966jv,
    author = "Hawking, Stephen",
    title = "{The Occurrence of singularities in cosmology. II}",
    doi = "10.1098/rspa.1966.0255",
    journal = "Proc. Roy. Soc. Lond. A",
    volume = "295",
    pages = "490--493",
    year = "1966"
}

@article{Hawking:1967ju,
    author = "Hawking, Stephen",
    title = "{The occurrence of singularities in cosmology. III. Causality and singularities}",
    doi = "10.1098/rspa.1967.0164",
    journal = "Proc. Roy. Soc. Lond. A",
    volume = "300",
    pages = "187--201",
    year = "1967"
}

@article{Bixano:2025bio,
    author = "Bixano, Leonel and Matos, Tonatiuh",
    title = "{On the Possibility of the Existence of Wormholes in Nature}",
    eprint = "2505.20167",
    archivePrefix = "arXiv",
    primaryClass = "gr-qc",
    month = "5",
    year = "2025"
}

@article{Bixano:2025qxp,
    author = "Bixano, Leonel and Matos, Tonatiuh",
    title = "{The space-time structure of an untouchable naked singularity}",
    eprint = "2508.01820",
    archivePrefix = "arXiv",
    primaryClass = "gr-qc",
    month = "8",
    year = "2025"
}

@article{Lobo:2002rp,
    author = "Lobo, Francisco and Crawford, Paulo",
    editor = "Buccheri, R. and Saniga, M. and Stuckey, W. M.",
    title = "{Time, closed time - like curves and causality}",
    eprint = "gr-qc/0206078",
    archivePrefix = "arXiv",
    journal = "NATO Sci. Ser. II",
    volume = "95",
    pages = "289--296",
    year = "2003"
}

@article{Minazzoli:2025gyw,
    author = "Minazzoli, Olivier and Wavasseur, Maxime and Chehab, Thomas",
    title = "{Deriving Entangled Relativity}",
    eprint = "2506.15209",
    archivePrefix = "arXiv",
    primaryClass = "gr-qc",
    month = "6",
    year = "2025"
}

@article{Minazzoli:2024zwo,
    author = "Minazzoli, Olivier",
    title = "{On the Principle of Relativity of Inertia in Both General and Entangled Relativities}",
    eprint = "2505.04667",
    archivePrefix = "arXiv",
    primaryClass = "gr-qc",
    doi = "10.1134/S1063779624701132",
    journal = "Phys. Part. Nucl.",
    volume = "55",
    number = "6",
    pages = "1488--1493",
    year = "2024"
}

@Article{Sarmiento-Alvarado2025,
author={Sarmiento-Alvarado, I. A.
and Wiederhold, P.
and Matos, T.},
title={Flat subspaces of the ${SL(n,\mathbb{R} )}$ chiral equations},
journal={General Relativity and Gravitation},
year={2025},
month={Sep},
day={19},
volume={57},
number={9},
pages={132},
issn={1572-9532},
doi={10.1007/s10714-025-03467-1},
adsurl={https://doi.org/10.1007/s10714-025-03467-1}
}

@Article{Sarmiento-Alvarado2023,
author={Sarmiento-Alvarado, I. A.
and Wiederhold, P.
and Matos, T.},
title={One-dimensional subspaces of the ${SL(n,\mathbb{R} )}$ chiral equations},
journal={International Journal of Theoretical Physics},
year={2023},
month={12},
day={23},
volume={62},
number={12},
pages={270},
issn={1572-9575},
doi={10.1007/s10773-023-05520-8},
adsurl={https://doi.org/10.1007/s10773-023-05520-8}
}

@article{LU2008511,
title = {Ricci-flat and charged wormholes in five dimensions},
journal = {Physics Letters B},
volume = {666},
number = {5},
pages = {511-516},
year = {2008},
issn = {0370-2693},
doi = {https://doi.org/10.1016/j.physletb.2008.07.100},
url = {https://www.sciencedirect.com/science/article/pii/S0370269308009763},
author = {H. Lü and Jianwei Mei}
}

@article{Dzhunushaliev:1998rz,
    author = "Dzhunushaliev, Vladimir D.",
    title = "{Multidimensional geometrical model of the renormalized electrical charge with splitting off the extra coordinates}",
    eprint = "gr-qc/9807080",
    archivePrefix = "arXiv",
    doi = "10.1142/S021773239800231X",
    journal = "Mod. Phys. Lett. A",
    volume = "13",
    pages = "2179--2186",
    year = "1998"
}

@article{Poisson:1995sv,
    author = "Poisson, Eric and Visser, Matt",
    title = "{Thin shell wormholes: Linearization stability}",
    eprint = "gr-qc/9506083",
    archivePrefix = "arXiv",
    doi = "10.1103/PhysRevD.52.7318",
    journal = "Phys. Rev. D",
    volume = "52",
    pages = "7318--7321",
    year = "1995"
}

@article{Lobo:2005us,
    author = "Lobo, Francisco S. N.",
    title = "{Phantom energy traversable wormholes}",
    eprint = "gr-qc/0502099",
    archivePrefix = "arXiv",
    doi = "10.1103/PhysRevD.71.084011",
    journal = "Phys. Rev. D",
    volume = "71",
    pages = "084011",
    year = "2005"
}

@article{Gao:2016bin,
    author = "Gao, Ping and Jafferis, Daniel Louis and Wall, Aron C.",
    title = "{Traversable Wormholes via a Double Trace Deformation}",
    eprint = "1608.05687",
    archivePrefix = "arXiv",
    primaryClass = "hep-th",
    doi = "10.1007/JHEP12(2017)151",
    journal = "JHEP",
    volume = "12",
    pages = "151",
    year = "2017"
}

@article{Maldacena:2018gjk,
    author = "Maldacena, Juan and Milekhin, Alexey and Popov, Fedor",
    title = "{Traversable wormholes in four dimensions}",
    eprint = "1807.04726",
    archivePrefix = "arXiv",
    primaryClass = "hep-th",
    doi = "10.1088/1361-6382/acde30",
    journal = "Class. Quant. Grav.",
    volume = "40",
    number = "15",
    pages = "155016",
    year = "2023"
}

@article{Ellis:1973yv,
    author = "Ellis, H. G.",
    title = "{Ether flow through a drainhole - a particle model in general relativity}",
    doi = "10.1063/1.1666161",
    journal = "J. Math. Phys.",
    volume = "14",
    pages = "104--118",
    year = "1973"
}

@article{Bronnikov:1973fh,
    author = "Bronnikov, K. A.",
    title = "{Scalar-tensor theory and scalar charge}",
    journal = "Acta Phys. Polon. B",
    volume = "4",
    pages = "251--266",
    year = "1973"
}

@article{Lazov:2017tjs,
    author = "Lazov, Boian and Nedkova, Petya and Yazadjiev, Stoytcho",
    title = "{Uniqueness theorem for static phantom wormholes in Einstein{\textendash}Maxwell-dilaton theory}",
    eprint = "1711.00290",
    archivePrefix = "arXiv",
    primaryClass = "gr-qc",
    doi = "10.1016/j.physletb.2018.01.059",
    journal = "Phys. Lett. B",
    volume = "778",
    pages = "408--413",
    year = "2018"
}

@article{Goulart:2017iko,
    author = "Goulart, Prieslei",
    title = "{Phantom wormholes in Einstein{\textendash}Maxwell-dilaton theory}",
    eprint = "1708.00935",
    archivePrefix = "arXiv",
    primaryClass = "gr-qc",
    doi = "10.1088/1361-6382/aa9dfc",
    journal = "Class. Quant. Grav.",
    volume = "35",
    number = "2",
    pages = "025012",
    year = "2018"
}

\end{document}